\documentclass[%
  aps,
  prapplied,
  preprint,             
  11pt,
  superscriptaddress,
  nofootinbib,
  floatfix,
  longbibliography,
  numbered              
]{revtex4-2}

\usepackage{amsmath,amssymb}
\usepackage{graphicx}
\usepackage{booktabs}
\usepackage{nicefrac}
\usepackage{float}
\usepackage{hyperref}
\usepackage{setspace}
\newcommand{\thickhline}{\hline\hline}

\begin{document}

\title{Machine learning inference of fission yields from gamma spectroscopy for very low-yield nuclear test verification}

\author{Julien de Troullioud de Lanversin}
\email[Contact author: ]{jdtdl@ust.hk}
\affiliation{Division of Public Policy, Academy of Interdisciplinary Studies, The Hong Kong University of Science and Technology, Hong Kong, China}

\author{Jiehui Li}
\affiliation{Division of Public Policy, Academy of Interdisciplinary Studies, The Hong Kong University of Science and Technology, Hong Kong, China}

\author{Christopher Fichtlscherer}
\affiliation{Department of Nuclear Science and Engineering, Massachusetts Institute of Technology, Cambridge, MA, U.S.}

\author{Dongdong She}
\affiliation{Department of Computer Science and Engineering, The Hong Kong University of Science and Technology, Hong Kong, China}

\author{Moritz K\"utt}
\affiliation{Department of Physics, University of Hamburg, Hamburg, Germany}

\date{April 27, 2026}

\begin{abstract}
Very low-yield nuclear tests pose a major verification challenge for the zero-yield standard of the Comprehensive Nuclear-Test-Ban Treaty (CTBT) and for existing testing moratoria. The zero-yield standard prohibits any explosive experiment that produces a self-sustaining fission chain reaction while allowing strictly subcritical experiments. Previous research shows that on-site gamma spectroscopy of post-test debris provides useful insight into the criticality level, although it remains heavily dependent on knowledge of certain experimental settings.
Here, we adopt a new approach whereby machine learning models are trained on simulated gamma spectroscopy data to infer the fission yield of a nuclear very low-yield test. Using high-fidelity 3-D Monte Carlo particle transport simulations, we generated gamma spectra measured outside containment vessels after very low-yield tests for 66 million representative scenarios. From these spectra, we extracted 82 fission-product-to-plutonium-239 peak ratios, then trained ML models for two tasks: (1) binary classification of whether a test exceeded a chosen yield threshold, and (2) regression to estimate the actual yield.
We find that the Extreme Gradient Boosting Method (XGBoost) performs best on the classification task among other popular methods across the most policy-relevant yield range (1\,g--100\,kg TNT). The classifier achieves high accuracy, even for yields near the chosen threshold (e.g., $>95\%$ for yields $\pm$ 100 g around a threshold at 1 kg TNT), and the regressor presents a mean absolute relative error of $12.4\%$, for measurements taken a month to a year after the test.
These results demonstrate that using machine learning to infer the yield of a past very low-yield nuclear test from gamma spectroscopy data is feasible and accurate. Combined with other verification methods and transparency measures, this approach can support efforts to establish a robust verification protocol for the zero-yield standard and could pave the way for a future yield threshold-based verification regime that is both technically feasible and politically viable.
\end{abstract}

\maketitle

\section{Introduction}
Nuclear-armed states have conducted nuclear weapon testing to develop new nuclear weapon designs, ensure their performance and safety, and signal technological capability. Due to concerns over environmental harm from radioactive pollution and to curb the nuclear arms race, states engaged in efforts to restrict these tests. In 1963, the Partial Test Ban Treaty (PTBT) entered into force and prohibited all nuclear tests in the atmosphere, underwater, and in outer space~\cite{USState1963}. The treaty did not restrict underground tests. In 1974, the Threshold Test Ban Treaty (TTBT) was negotiated~\cite{USStateTTBT2009}. This treaty between the U.S. and the USSR bans non-peaceful underground tests with a yield exceeding 150 kilotons TNT equivalent (shortened to ``TNT'' in subsequent instances). To prevent this limit from being circumvented under the guise of civil engineering projects, the Peaceful Nuclear Explosions Treaty (PNET) was signed in 1976, applying the same 150-kiloton cap to non-military explosions. The last major step toward banning nuclear weapon testing happened with the negotiations of the Comprehensive Nuclear-Test-Ban Treaty (CTBT), which opened for signature in 1996~\cite{ctbt}. The CTBT prohibits ``any nuclear weapon test explosion or any other nuclear explosion'' by any state, in any physical environment. The treaty has not yet entered into force, as key states, including the U.S., China, and Russia, have not ratified it.\footnote{Russia withdrew its 1996 ratification in 2023.}

Yet, since 1998, all nuclear-armed states except North Korea have voluntarily declared moratoria on nuclear weapons testing, aligning with the CTBT's provisions. The Preparatory Commission for the Comprehensive Nuclear-Test-Ban Treaty Organization (CTBTO) is tasked with establishing the verification regime through its International Monitoring System (IMS). This system employs a global network of seismic, hydroacoustic, infrasound, and radionuclide monitoring stations to detect nuclear explosions exceeding 100 tons of TNT worldwide and at lower yields in specific regions~\cite{NAS-2002}.

During the CTBT negotiations, some states and civil groups pushed for the treaty to be a strict ``zero-yield'' treaty where any explosive test or experiment involving fission reactions would be prohibited. 
By banning all nuclear explosive tests, the treaty would aim to constrain the advancement of nuclear arsenals and limit the proliferation of nuclear weapons capabilities. 
However, U.S. negotiators and experts nuanced the interpretation of zero-yield, remarking that technically any inert mass of fissile material had a non-zero fission yield~\cite{SAFETY_AND_RELIABILITY_OF_THE_US}. The U.S. proposed that in the context of the CTBT, ``zero-yield'' meant that any nuclear explosive test involving a self-sustaining fission chain reaction should be prohibited. This interpretation, known as the ``zero-yield standard,'' has been largely followed by other nuclear-armed states~\cite{P5-Public-Statements}. This standard technically allows states to conduct explosive subcritical experiments as long as they do not generate exponentially growing or self-sustaining fission chain reactions, regardless of the yield~\cite{zero-yield-standard}.

In order to maintain confidence in the performance and safety of their nuclear arsenal without relying on full-scale nuclear tests, nuclear-armed states have been conducting various laboratory-scale experiments involving weapons-grade fissile materials, sometimes in conditions nearing those of a real nuclear weapon detonation~\cite{Garwin-Simonenko-1996}. One category of such experiments comprises very low-yield tests, which involve compressing weapons-grade fissile material (usually weapons-grade plutonium) with conventional explosives to initiate a fission chain reaction that yields only a very limited amount of energy. These tests are conducted in underground chambers, usually within a containment vessel that can withstand the yield. Very low-yield tests that only involve a subcritical fission chain reaction and thus comply with the zero-yield standard are called subcritical tests. The U.S., Russia, and China are currently the only states allegedly conducting subcritical tests~\cite{Lanversin-Fichtlscherer-2024}.

In recent years, growing concerns have emerged regarding potential non-compliance with the zero-yield standard by certain nuclear-weapon states. In 2019, the U.S. government alleged that Russia, and possibly China, had conducted very low-yield supercritical tests. In October 2025, U.S. President Donald J. Trump ``instructed'' his administration to ``start testing our Nuclear Weapons on an equal basis," likely referring to non-compliance at very low yields the U.S. said it observed in China and Russia~\cite{kimball2025}. In February 2026, the U.S. government provided more details on its accusation of non-compliance against China~\cite{stone2026}. These allegations have prompted discussions among U.S. officials about resuming full-scale nuclear testing and resurfaced a key concern among critics of CTBT ratification in the U.S. Congress: the absence of robust verification mechanisms to detect very low-yield tests and ensure adherence to the zero-yield standard in the current IMS~\cite{WashingtonPost-2020, McGrath-2009}. In 2023, the U.S. administration highlighted the need to develop on-site procedures to verify that experiments remain subcritical~\cite{hruby2023}.

The authors' ongoing research demonstrates that on-site gamma spectroscopy of the radioactive debris remaining in the vessel after a very low-yield test, together with the activation products in the vessel's wall, can provide insights into the test's yield and its criticality level. However, the results indicate that this information is limited for two primary reasons: (1) unlike the yield, the criticality level does not produce a direct signature in the gamma spectrum from the radioactive debris, and (2) inferences about the yield and the criticality level are confounded by other factors influencing the measured gamma spectrum, which can remain unknown to inspectors.

To address these difficulties, the present work explores how machine learning (ML) methods can be applied to these gamma signatures to infer information about a very low-yield test. We generated 66 million gamma spectra using high-fidelity models of very low-yield tests across a wide range of test and measurement parameters. Using ratios of gamma lines as input data, the Gradient Boosting method can classify and estimate the test yields with high accuracy, even months to a year after the test. 

This research demonstrates that post-test verification of very low-yield tests using gamma spectroscopy is feasible and reliable for yield inference. A verification protocol based on this method could strengthen adherence to the current moratoria and help facilitate the entry into force of the CTBT.
Yet, the current zero-yield standard focuses on the criticality level of an experiment, but it is not explicitly defined in the CTBT, and its interpretation is not formally codified. 
History shows that identifying such ambiguities has frequently catalyzed refinements and enhancements to arms control agreements. Thus, this technical research could spark productive discussions on the interpretation of the zero-yield standard, the adequacy of this interpretation with respect to verification feasibility, and, ultimately, the establishment of a verification regime that secures the confidence of all CTBT States Signatories.

\section{Physics of very low-yield tests and post-test gamma spectroscopy}

\subsection{Yield and criticality level of a very low-yield test}

The criticality level of a fission chain reaction happening in a plutonium assembly is often represented by $\alpha(t)$, the time-dependent prompt neutron decay constant. This parameter drives the change in fission neutron population over time by:


\begin{eqnarray}
\label{eq:alpha}
\frac{dN(t)}{dt} = \alpha(t)N(t)
\end{eqnarray}

where $N(t)$ is the neutron population at time $t$~\cite{primer}. The fission chain reaction is said to be supercritical when $\alpha(t)$ is positive and subcritical when $\alpha(t)$ is negative. This parameter depends on the assembly's composition, mass, and density, as well as the presence of a reflector surrounding it. For a given mass, $\alpha(t)$ will increase when the assembly is compressed. During a very low-yield test, $\alpha(t)$ will first increase as the assembly is compressed by high explosives and then decrease as it expands back. Details on the time evolution of $\alpha(t)$ during a very low-yield test are rarely accessible in unclassified literature. In a supercritical test, its value will briefly become positive when the assembly is at its maximum compression, whereas it always remains below zero in a subcritical test.

Because the number of fission events over time is proportional to the neutron population, the total fission yield of a very low-yield test can be approximated as follows:

\begin{eqnarray}
\label{eq:yield}
Y =  E_{\text{f}}N_{F,0} \int_{0}^{T} \text{e}^{\int_0^{t}\alpha(t') dt'} \text{d}t 
\end{eqnarray}
where $E_{\text{f}}$ is the energy released per fission event, $N_{F,0}$ is the initial number of fissions created by initiating neutrons or high-energy photons, and $T$ is the time after which the assembly has expanded to a degree such that the reaction can be considered to have effectively stopped. This equation shows that the yield of a test grows exponentially with $\alpha(t)$.

\subsection{Post-test gamma spectroscopy of very low-yield tests}

\begin{figure}[H]
\centering
\includegraphics[width=0.7\linewidth]{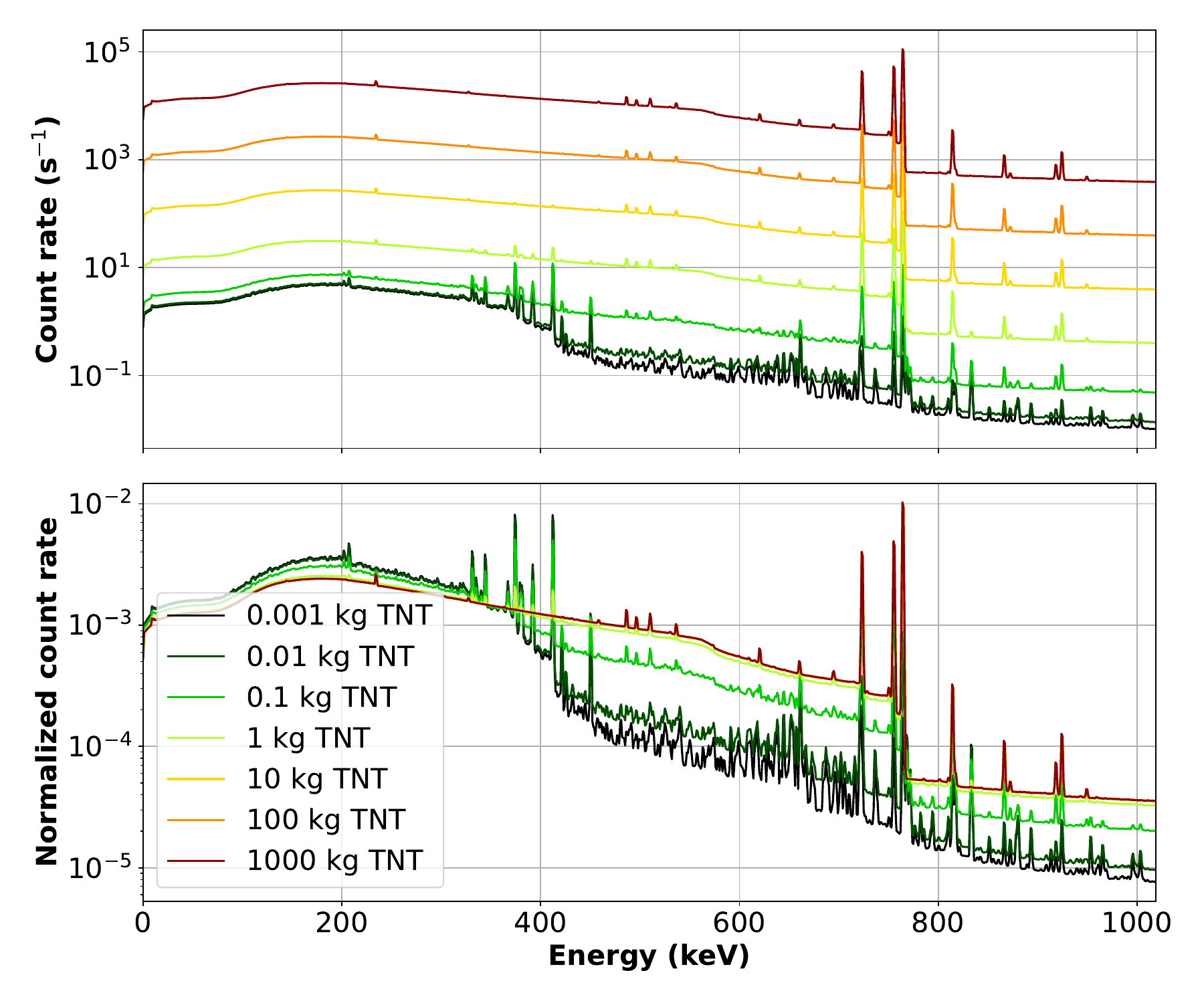}
\caption{{\bf Gamma spectra for different yields measured one month after a test.} The top graph presents counts per second, while the bottom graph presents count rates normalized by the total count across all energies. The spectra were obtained by simulating gamma detection with a simulated HPGe detector.}
\label{fig:spectra}
\end{figure}

The debris within the containment vessel after a test comprises residual plutonium, fission products, and other materials originally present in the vessel, such as diagnostic components or structural elements. The radioactive fission products emit characteristic gamma rays that propagate through the system, some of which escape outside the vessel. Figure~\ref{fig:spectra} illustrates simulated gamma spectra that would be measured one month after tests of various yields, as detected outside the vessel. As shown in the graphs, the measured gamma spectra outside the vessel strongly depend on the test yield, which determines the quantity of fission products generated. The spectra's dependence on $\alpha(t)$ is fully captured within the yield itself, with no other observable indicators in the spectroscopy measurements that reveal the sign of $\alpha(t)$. Another work by the same authors shows how the sign of $\alpha(t)$ can be inferred under specific conditions, based on the ranges of other test parameters~\cite{PRApplied}. The present work, however, focuses exclusively on determining the yield from the measured spectra.

Other factors besides the yield influence the gamma spectra measured outside the vessel. These include: the distribution of debris inside the vessel, the mass and configuration of the tested assembly before the test, the shielding effects as gamma rays traverse materials, and the time elapsed between the test and the measurements.

The dependence on the spatial distribution of the debris can be mitigated by calculating ratios of gamma lines emitted by a fission product over lines emitted by the isotope plutonium-239, assuming all fission product species are similarly distributed in the vessel. With the approximation that plutonium-239 fissions dominate, the ratio for a given line $l$ of a fission product $i$ and for a given line $l_{\text{Pu}}$ from plutonium-239 can be expressed as:\footnote{Details on the derivation of this equation can be found in~\cite{PRApplied}}

\begin{eqnarray}
\label{eq:ratio}
R_{\text{i}, \text{l}, \text{l}_{\text{Pu}}}(Y) = \bigg[ \frac{\lambda_{\text{i}} I_{\text{l}}}{\lambda_{\text{Pu}} I_{\text{l}_{\text{Pu}}}} \bigg] \cdot  S(\text{l}, \text{l}_{\text{Pu}}),  \cdot D_{\text{i}}(\Delta t) \cdot \bigg[ r_{\text{i,pre}} + r_{\text{i,test}}(Y, m_{\text{Pu}})\bigg].
\end{eqnarray}

Terms in the first bracket are known nuclear decay data with the decay constants $\lambda$ for the fission product $i$ and plutonium-239 and gamma emission line intensities $I$ for selected lines.
The term $S$, which summarizes the shielding effects, is equal to the ratio of the transmittance of the fission product's gamma line $l$ over the plutonium-239 line $l_{\text{Pu}}$, i.e. $S = \nicefrac{T_{\text{l}}}{T_{\text{l}_{\text{Pu}}}}$. The term $D_{\text{i}}$ accounts for the time evolution of fission product $i$ over the time interval $\Delta t$ between the test and the measurement (referred thereafter as the time after test). 

The term $r_{\text{i,pre}}$ is the ratio of fission product $i$ created by spontaneous and induced fissions in the assembly before the experiment over plutonium-239. It is mainly affected by $\Delta t_{\text{sep}}$, which measures the period from plutonium separation until the test, and by $\alpha_{\text{pre}}$, the criticality level of the assembly's configuration over that period.
For the majority of fission products, their concentrations move toward an equilibrium over time. 
The present work assumes a gap of at least one year between plutonium separation and testing. By limiting the analysis to fission products that achieve equilibrium within one year, $r_{\text{i,pre}}$ is determined only by $\alpha_{\text{pre}}$.

The term $r_{\text{i,test}}$ is the ratio of fission product $i$ generated during the test when the assembly is irradiated over plutonium-239. This term relates to the yield $Y$ as follows:

\begin{eqnarray}
\label{eq:ratio_pre}
r_{\text{i,test}}(Y, m_{\text{Pu}}) =   FY_{\text{i}}\frac{M_\text{Pu}}{m_{\text{Pu}}} \frac{1}{E_{\text{f}}} Y
\end{eqnarray}
where $FY_{\text{i}}$ is the cumulative fission yield of fission product $i$ from plutonium-239, $M_{\text{Pu}}$ is the plutonium-239 atomic mass and $m_{\text{Pu}}$ the mass of plutonium-239 used in the test.

Equations \ref{eq:ratio} and \ref{eq:ratio_pre} show that, under certain approximations, the ratio $ R_{\text{i}, \text{l}, \text{l}_{\text{Pu}}}(Y) $ exhibits a linear relationship with the yield $ Y $. In practice, however, inferring the yield from measured ratios is challenging for three primary reasons: 1) parameters such as $ S $, $ \Delta t $, $\alpha_{\text{pre}}$, and $ m_{\text{Pu}} $ are unknown to inspectors and cannot always be independently
verified; 2) triangulation by using multiple ratios is complicated by the dependence of
some of these parameters on the specific fission product $i$; 3) measured gamma spectra
are subject to statistical and instrumental uncertainties, as well as background noise.

\section{Machine learning methods for yield inference}

\subsection{Machine learning applications in gamma-ray spectroscopy}

ML has become a powerful tool for extracting quantitative information from gamma-ray spectra, particularly when complex spectral overlaps, background noise, shielding effects, and unknown confounding parameters limit traditional analytical approaches \cite{zehtabvar2024}. In nuclear safeguards and security applications, ML algorithms have been applied to automated radionuclide identification, peak deconvolution, activity estimation, and quantitative characterization of nuclear material, even with low-resolution detectors, low-count statistics, or heavily shielded conditions \cite{bandstra2023,galib2021,Kirchknopf_2024}. Recent systematic reviews confirm the rapid progress in this area and highlight the strengths of both deep learning
and traditional methods for these tasks \cite{zehtabvar2024, sun2025}. Similar ML-based approaches have also been used in inertial-confinement fusion diagnostics, where models trained on simulated spectra infer reaction yields from gamma-ray measurements \cite{Landsmeer_2025}.

Gradient-boosted decision trees (GBDT) have emerged as especially effective in these domains because of their robustness to noisy spectroscopic data and inherent feature-importance ranking \cite{Romo_2021}. GBDT models have been shown to perform well when trained on realistic (often Monte Carlo-generated) gamma-ray spectra, even under conditions where traditional analytical methods struggle \cite{galib2021}. These precedents motivated the present work, which applies ML methods,
and GBDT in particular, to the novel task of inferring fission yield from post-test gamma spectra of very low-yield nuclear experiments.

\subsection{Classification and regression tasks}

This work investigates the feasibility and performance of ML methods for two tasks: classifying the yield below or above a given threshold (classification task), and estimating the value of the yield (regression task). The classification task relates to possible agreed-upon yield thresholds as proposed in the 1974 TTBT and as discussed in the CTBT negotiations before the zero-yield standard prevailed. Under this approach, all that is required is to verify whether the test's yield was above or below a certain threshold. The second relates to a context in which the actual value of the yield is necessary to verify a very low-yield test. For instance, estimating the yield value can significantly improve knowledge of the criticality level of a test, or its possible application for weapon development. Estimating the actual yield can also be used to classify a test against a yield threshold, but it provides more information than necessary and can complicate the verification process if the host considers the actual yield value sensitive. 

\section{Computational models}

\begin{figure}[H]
\centering
\includegraphics[width=1.0\linewidth]{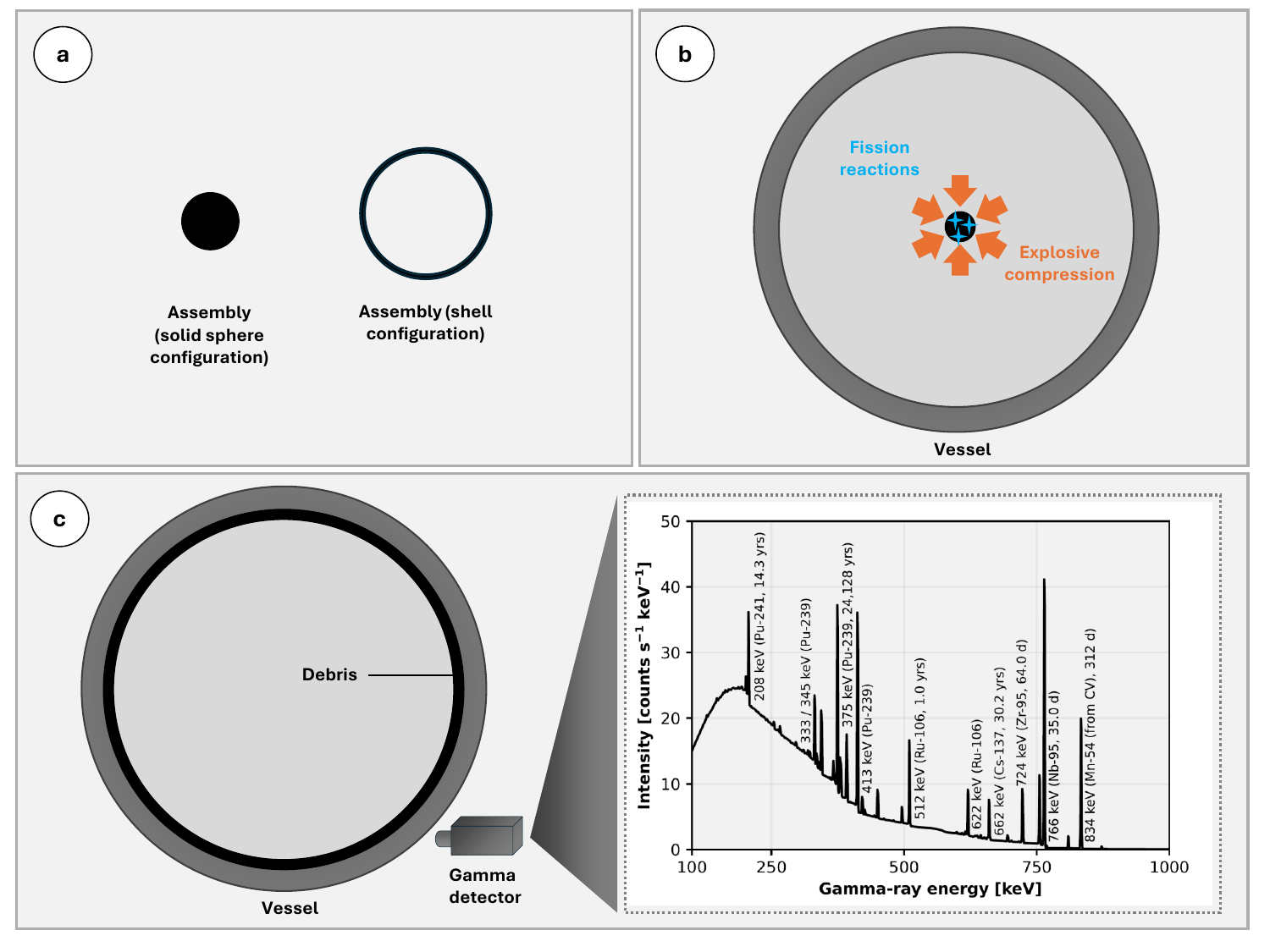}
\caption{{\bf Schematic configurations of the models used for simulations.}
 a. illustrates the two pre-test configurations of the assembly that were used to model fission product equilibrium before the test. b. represents the configuration used during the test where the assembly is explosively compressed and irradiated by fission-initiating neutrons. c. represents the post-test configuration used to generate measured gamma spectra where the radioactive debris is homogeneously distributed on the inner wall of the vessel and the gamma detector is placed 20 cm from the vessel.}
\label{fig:geo}
\end{figure}

To generate measured spectra, we simulated very low-yield tests, including physical processes before, during, and after the test, and gamma measurements with a High Purity Germanium (HPGe) radiation detector, using 3-dimensional computer models. Figure \ref{fig:geo} a--c presents the geometrical configuration of these models.

\subsection{Pre-test model}

As shown in Figure \ref{fig:geo} a., two pre-test configurations are modeled for the assembly: a solid sphere and a shell. Both are made of the same weapons-grade plutonium.\footnote{The composition for the weapons-grade plutonium is as follows: plutonium-238: 0.012~\%; plutonium-239: 93.81\%; plutonium-240: 5.81\%; plutonium-241: 0.349\%; plutonium-242: 0.022\% \cite{Nicholas-LANL}} For each configuration, the accumulation of fission products via spontaneous and induced fissions in the assembly is modeled with the coupling of the open-source particle transport code OpenMC and an in-house module named \textit{decaypy} (\url{https://github.com/cfichtlscherer/decaypy}) \cite{OpenMC}. This routine involves a three-phase procedure: Initially, using information from the Fission Reaction Event Yield Algorithm (FREYA), we calculated the production rate and energy spectrum of neutrons arising from spontaneous fission~\cite{FREYA}. These neutrons served as the input source for fixed-source simulations conducted in OpenMC, enabling the computation of induced fission rates.\footnote{These rates depend on the assembly's criticality configuration in the years preceding the test.} In the concluding phase, the rates of these induced fissions were integrated into the decay modeling by deriving effective pseudo-half-lives to represent the fission contributions alongside the standard decay.

\subsection{Very low-yield test model}

This model is used for simulating the actual test (Figure \ref{fig:geo} b.). The modeled containment vessel is made of an alloyed stainless-steel which composition is taken from Table A1 in a 2012 NAVSEA technical publication~\cite{NAVSEA2012}. The initial composition of the compressed assembly (i.e., plutonium and fission products) depends on its pre-test configuration.

In this phase, OpenMC is coupled with the open-source depletion code ONIX to compute the neutronics and isotopic changes in the assembly and the vessel material over time~\cite{ONIX}. Beginning at time $t = 0$, using the initial isotopic composition supplied by decaypy, an OpenMC simulation runs $ 10^{6} $ neutron histories to determine the neutron reaction rates and flux within the compressed assembly and the vessel. Before being fed to ONIX, the flux is adjusted based on the power produced in the assembly through fission processes. ONIX then solves the associated depletion equations to derive the isotopic compositions in the assembly and in the vessel at time $t = T$, when the test is considered finished.

\subsection{Post-test debris, gamma radiation, and gamma detector models}

After the test, the debris resulting from the disintegration of the assembly is assumed to be evenly distributed on the inner surface of the vessel wall (Figure \ref{fig:geo} c.).\footnote{Other distributions of the debris inside the vessel have been simulated, and the resulting gamma emission signature did not significantly deviate from those associated with a uniform distribution.} ONIX was used to simulate the decay of radioactive elements over time and compute isotopic changes months to a year after the test.

Gamma emissions originate from both radioactive elements in the debris and activation products in the vessel. The energies and intensities of gamma rays emitted by the radioactive isotopes were established based on the data from the Evaluated Nuclear Structured Data Files (ENSDF) supplied by the National Nuclear Data Center (NNDC)~\cite{NuDatNationalNuclear}.
Photon emission currents from the vessel were derived by modeling the transport of these gamma rays.
This simulation relied on OpenMC's verified photon transport features \cite{lundImplementationValidationPhoton2018}.

To generate realistic measured gamma spectra, a HPGe detector was modeled, positioned 20 cm from the vessel's surface, and featuring a 38.5 cm$^2$ active area.
The detector's response was calculated using the validated pulse-height tally method combined with Gaussian energy broadening, where the resolution was derived from assessments of real-world experimental data \cite{fichtlschererModelingGammaDetectors2024, goodellInvestigatingPracticalityMinimally2019}. More information on the detector modeling and its response can be found in \cite{PRApplied}.
By convolving the detector response function with the calculated photon current, the simulated measurements from the detector were obtained.

\section{Data generation, training procedure, and evaluation methods}

\subsection{Data generation}

We generated a total of 66,000,000 spectral datasets using the previously described computational models, varying parameters such as yield, time after test, shielding effects (which are varied with the thickness of the vessel in cm, $t_{\text{vessel}}$),\footnote{In a real-world scenario, shielding effects will be caused by a combination of self-shielding, vessel shielding, and shielding from other materials inside and outside the vessel through which gamma rays pass.} plutonium mass, pre-test criticality configuration, measurement time, and finally by adding statistical noise. Each spectrum comprises 3,000 1-keV-wide bins. Table~\ref{tab:parameters} lists the ranges taken and the method of sampling for each parameter.

\begin{table}[!ht]
\caption{\textbf{Information on parameters' ranges}. The fourth column specifies the method used to sample values within the specified range.}
\begin{tabular}{l|l|l|l}
\textbf{Parameter} & \textbf{Range} & \textbf{Sampling size   } & \textbf{Sampling method} \\
\thickhline
\textbf{Unknown parameters} &  &  & \\
$Y$ & 0.1 mg -- 1000 kg TNT & 1000 & Logarithmic \\
$\Delta t$ & 30 -- 365 days & 11 & Linear \\ 
$t_{\text{vessel}}$ & 3 -- 8 cm & 6 & Linear \\ 
$m_{\text{Pu}}$ & 0.15 -- 3 kg & 10 & Linear \\
$\alpha_{\text{pre}}$ & Shell -- Solid sphere & 2 & --- \\
\hline
\textbf{Controlled parameter} &  &  & \\
$t_m$ & 1.5 -- 24 hours & 5 & Logarithmic \\ 
\end{tabular}
\label{tab:parameters}
\end{table}

The measurement time of the simulated gamma spectra is not considered as a confounding parameter in this work, as the inspectors would have control over it during a verification process. When not otherwise specified, all results presented in this work were obtained for spectra measured for 24 hours.


Statistical noise was added by sampling the total number of photons $N$ from a Poisson distribution $\mathrm{Poisson}(C_{\mathrm{tot}})$ (where $C_{\mathrm{tot}}$ is the computed total count for a given spectrum), then drawing the noisy spectrum from $\mathrm{Multinomial}(N, p)$, where $p_i = s_i / C_{\mathrm{tot}}$. This accurately models Poisson fluctuations in gamma-ray photon counting. For each simulated spectrum, ten noisy variants were created.

\subsection{Yield range selection}

As shown in Table \ref{tab:parameters}, we initially generated spectral data and explored results for yields ranging from 0.1~mg to 1~ton TNT. While 0.1~mg represents a typical yield for a subcritical test, 1 ton is often cited as the yield limit under which a test becomes invisible for the IMS~\cite{NRC2012}. In this paper, we narrow down our focus to yields ranging from 1 g to 100 kg TNT for the following reasons. According to previous work conducted by the same authors, yields of a few tens of grams mark the limit above which a test can be unambiguously characterized as being supercritical (subcritical tests cannot produce yields above that limit). This yield region is therefore relevant for the current zero-yield standard. This range also includes previously proposed yield thresholds discussed during the CTBT negotiations.\footnote{Before the zero-yield standard was accepted, the U.S. proposed a 1.8 kg TNT threshold and the U.K. proposed a 45 kg threshold ~\cite{Johnson2009}.} Finally, it stops at the upper yield limit of hydronuclear tests\footnote{Hydronuclear tests are very low-yield supercritical tests that have historically been used by the U.S. and the USSR to
investigate the safety of their nuclear weapon designs and assess the
performance of new designs.} as defined by the USSR~\cite{NRC2012}
We consider the range from 1 gram to 100 kg TNT to be the most relevant yield window where verification processes should assess compliance with the current zero-yield standard and where possible yield threshold limit might be located.

\subsection{Data processing and model training}

Eighty-two\footnote{The selection of these ratios is based on \cite{PRApplied}} fission product lines were selected to calculate peak ratios with the plutonium-239 line at 375.1~keV from each simulated gamma spectrum. To ensure physical consistency with radiation physics principles and to maximize the interpretability of the model predictions (essential in a nuclear verification context where decisions on potential non-compliance must be transparent and defensible) we adopted a physics-informed approach for data processing and model training.

Consistent with the statistical noise and limited energy resolution of the simulated spectra, a detectability limit was established for all gamma peaks based on the Currie equation \cite{Knoll}.\footnote{Detectability was defined using the standard Currie detection limit, corresponding to the conventional one-sided ($\alpha=\beta=0.05$) criterion (approximately 95\% confidence for both false-positive control and detection power).} Spectra in which the plutonium-239 reference line could not be detected were excluded from the dataset. For the retained spectra, peak ratios corresponding to undetectable fission-product lines were set to zero. 

The final dataset used for training and evaluating the ML models therefore consisted of all computed peak ratios (with undetectable ratios set to zero), together with their associated test and measurement parameters, from the 66,000,000 raw spectra in which the plutonium line was detectable. The number of retained spectra and the number of detectable fission-product lines vary significantly with yield and other parameters; further details are provided in the Appendix \ref{app:S1_Figure}. 

Five-fold stratified cross-validation was performed using \texttt{StratifiedGroupKFold} from scikit-learn. Groups were defined by the original selected spectra so that the ten noisy variants generated from each spectrum were kept together in the same fold. This grouping strategy prevents data leakage between the training and test sets while preserving class proportions across folds. Hyperparameters of all six models tested were optimized via randomized search cross-validation using \texttt{RandomizedSearchCV} from scikit-learn.

\subsection{Evaluation and Analysis}

This work uses different sets of metrics to investigate and showcase the trained model's performance on classification and regression tasks.

For the classification task, a positive prediction occurs when the model classifies an instance as exceeding the yield threshold (i.e., a violation), while a negative prediction occurs when it classifies an instance as falling below the threshold (i.e., compliance). 
Five standard performance metrics (misclassification rate, accuracy, precision, recall, and F1 score) were computed to evaluate the models. 

For the regression task, in which the model predicts the actual yield, three main performance metrics are used: relative error, mean absolute error in logarithmic space (log-MAE), and the percentage of predictions within a factor of \(x\). 
The mean absolute error in logarithmic units is the average of the absolute differences between the \(\log_{10}\)-transformed predicted and actual yields; this transformation ensures that errors at low and high yields receive equal weight. 

SHAP (SHapley Additive exPlanations) analysis was performed for both the classification and regression tasks to identify and rank the peak ratios that the models rely on most heavily for making predictions. We used TreeSHAP on each fold’s held-out validation data in 5-fold grouped cross-validation, and report feature importance as mean absolute SHAP values aggregated across all out-of-fold samples. 

The Results section further examines how variations in the key test and measurement factors (i.e., time between testing and measurement, shielding effect, and plutonium mass) and different training strategies (such as changing the training ranges of these factors or adding them as input features alongside the peak ratios) affect overall model performance.

\section{Results}

\subsection{Machine learning models comparison}

\begin{table}[!ht]
\caption{\textbf{Performance for different ML methods.} Comparison of F1 score for six classification models at six yield thresholds. The complete yield range (0.1 mg to 1 ton TNT) was used for the training and testing window for each model and each threshold. Bold indicates the column-wise maximum. 
\textbf{XGBoost} = eXtreme Gradient Boosting, 
\textbf{RF} = Random Forest, 
\textbf{DT} = Decision Tree, 
\textbf{MLP} = Multi-Layer Perceptron, 
\textbf{KNN} = K-Nearest Neighbors, 
\textbf{SVC} = Support Vector Classifier.}
\begin{tabular}{l|l|l|l|l|l|l}
\textbf{Model} & \textbf{0.001 kg TNT} & \textbf{0.01 kg TNT} & \textbf{0.1 kg TNT} & \textbf{1 kg TNT} & \textbf{10 kg TNT} & \textbf{100 kg TNT}\\
\thickhline
XGBoost & \textbf{0.9958} & \textbf{0.9972} & \textbf{0.9985} & \textbf{0.9974} & \textbf{0.9938} & \textbf{0.9830} \\
RF & 0.9953 & 0.9968 & 0.9984 & 0.9970 & 0.9930 & 0.9823 \\
DT & 0.9888 & 0.9926 & 0.9973 & 0.9949 & 0.9897 & 0.9746 \\
MLP & 0.9027 & 0.9646 & 0.9900 & 0.9928 & 0.9866 & 0.9691 \\
KNN & 0.8923 & 0.9568 & 0.9937 & 0.9944 & 0.9843 & 0.9540 \\
SVC & 0.8822 & 0.9359 & 0.9816 & 0.9747 & 0.9515 & 0.9083 \\
\end{tabular}
\label{tab:model-metrics-binary-classification}
\end{table}

Table~\ref{tab:model-metrics-binary-classification} presents the F1 scores of six selected ML models for the classification task at different yield thresholds. Details on the models' architectures and specific settings are provided in Appendix \ref{app:S1_Text}. Here, the complete yield range (0.1 mg to 1 ton TNT) was used when training and testing each classification model for each threshold. As a result, the F1 score is inflated but still relevant for comparing models' performance. Tree-based methods (XGBoost, Random Forest, and Decision Tree) have the highest score at each threshold, with XGBoost having a slight edge over the other two. Multi-layer perceptrons (MLP), the representative for deep learning methods, fall slightly behind gradient-boosted tree ensembles in terms of performance. Other methods such as KNN (K-nearest neighbor) and SVC (Support Vector Classifier) struggle with this task to a different degree. XGBoost is the ML method selected in this work to produce subsequent results and analysis.



\subsection{General performances}

\subsubsection{Classification}

\begin{figure}[H]
\centering
\includegraphics[width=0.8\linewidth]{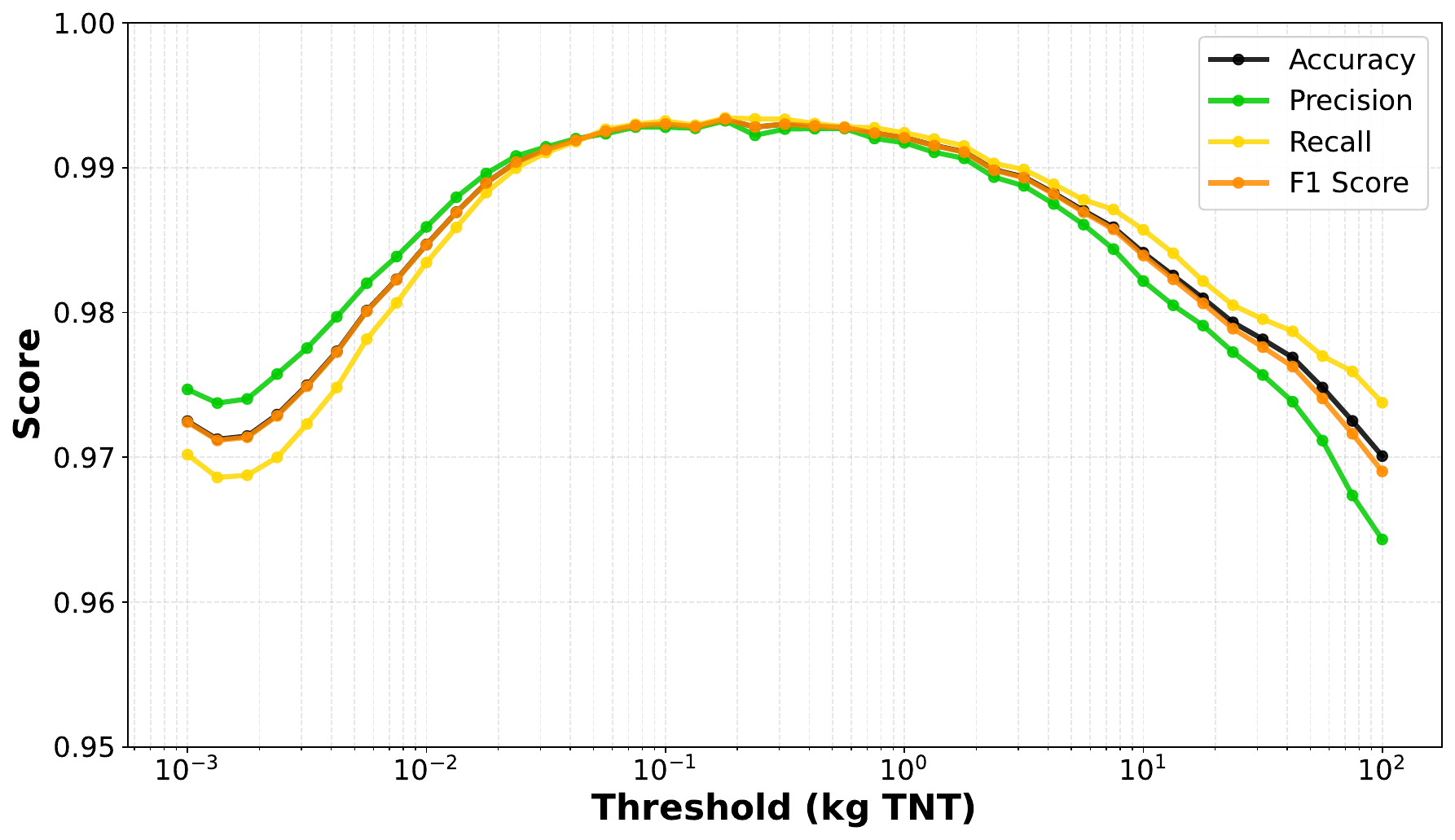}
\caption{{\bf Metrics' scores across yield thresholds for the classification task.} Metrics were computed for models trained and tested in a range of one order of magnitude above and below each threshold.}
\label{fig:classification_metrics}
\end{figure}

Figure ~\ref{fig:classification_metrics} plots metrics for the classification task across a range of thresholds from 1 g to 100 kg TNT. For each threshold, the model was only trained and tested on a local yield window spanning one order of magnitude on both sides of the threshold (e.g. from 100 g to 10 kg for a threshold at 1 kg TNT).\footnote{Using a local range for training and testing prevents scores inflation whereby instances with extreme yield values are easy to classify and artificially increase the score}  As seen on the graph, all metrics score above 0.96 across thresholds with a plateau from 100 g to 1 kg TNT. The performance gradually drops toward both edges of the yield range. As will be seen in subsequent results, this pattern is also encountered for the regression model and stems from the drop in data availability and quality for lower and higher yield regions as described in Appendix \ref{app:S1_Figure}.

\begin{figure}[H]
\centering
\includegraphics[width=1.0\linewidth]{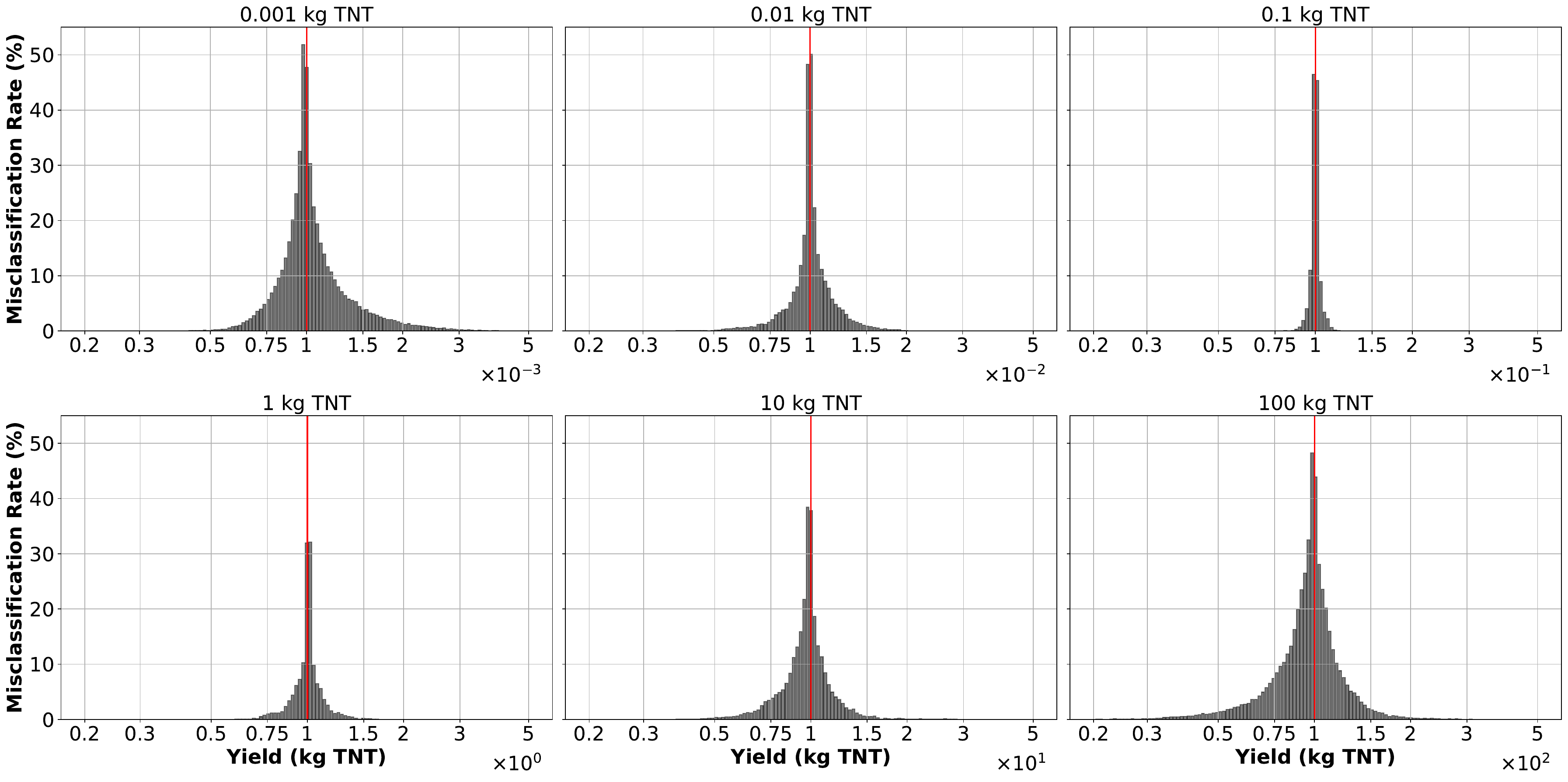}
\caption{{\bf Misclassification rate for various yield thresholds.}
 Each graph shows the distribution of misclassification rates for groups of same-yield instances around a specific yield threshold. The red line indicates the yield threshold. Each same-yield group comprises 13,200 instances.}
\label{fig:misclassification_yield}
\end{figure}

Figure \ref{fig:misclassification_yield} shows the distribution of misclassification rates for selected yield thresholds at 0.001, 0.01, 0.1, 1, 10, and 100 kg TNT. The training and evaluation of the models were conducted on local threshold-specific windows as described previously. In all six plots, misclassifications are highly concentrated around the yield threshold (indicated by the vertical red line) and rapidly decrease to near zero as the yield values move away from the threshold, indicating that almost all errors occur for values close to the decision boundary.

These graphs are helpful to visualize the risks of false positives and negatives in the context of a verification. Taking the 10 kg TNT threshold as an example, the model's misclassification rate depends heavily on the true yield of the event. For an actual yield of 13 kg TNT, the system exhibits a false negative rate of 2\%, meaning 2\% of such events would incorrectly be classified as compliant. Conversely, for an actual yield of 9 kg TNT, the system has a false positive rate of 13\%, meaning 13\% of these compliant events would be incorrectly flagged as violations.


Figure \ref{fig:classification_metrics} and Figure \ref{fig:misclassification_yield} show that the model for the classification task performs best for thresholds around 100 g TNT.

\subsubsection{Regression}

\begin{figure}[H]
\centering
\includegraphics[width=0.8\linewidth]{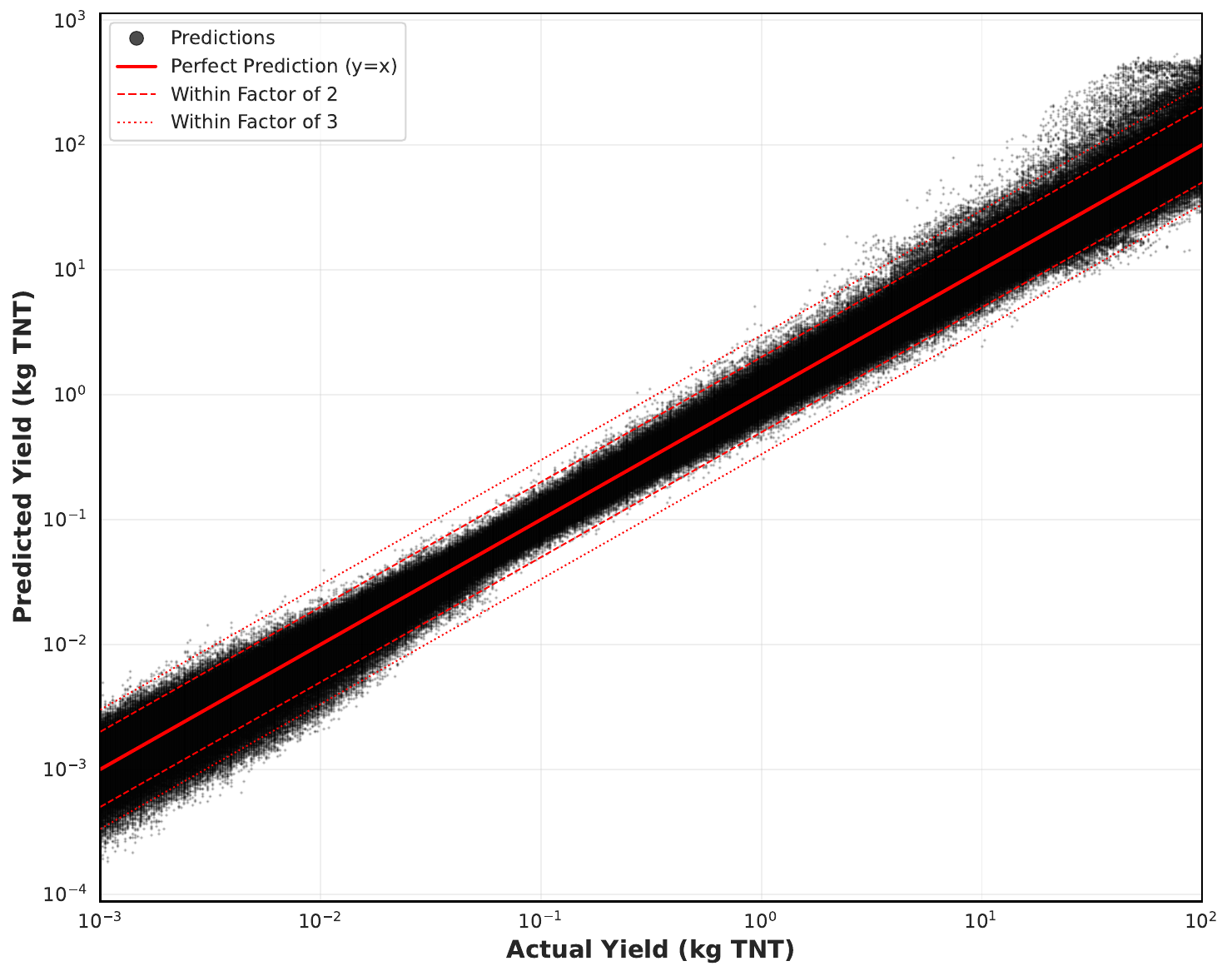}
\caption{{\bf Predicted vs. actual yield values.} Each prediction is plotted as a black dot. The red plain line represents perfect prediction. The red dashed and dotted lines delimit regions where the prediction is within a factor of 2 or 3 of the actual value.}
\label{fig:regression_predicted_actual}
\end{figure}

Figure \ref{fig:regression_predicted_actual} shows the regression model's predicted yield against the actual yield for each dataset instance from 1 g to 100 kg TNT. We observe that predictions are overall unbiased, as the cloud of predicted points is evenly distributed around the line of perfect prediction. We also note that the model is less precise in predicting yield in the lower and higher ranges.

\begin{figure}[H]
\centering
\includegraphics[width=0.8\linewidth]{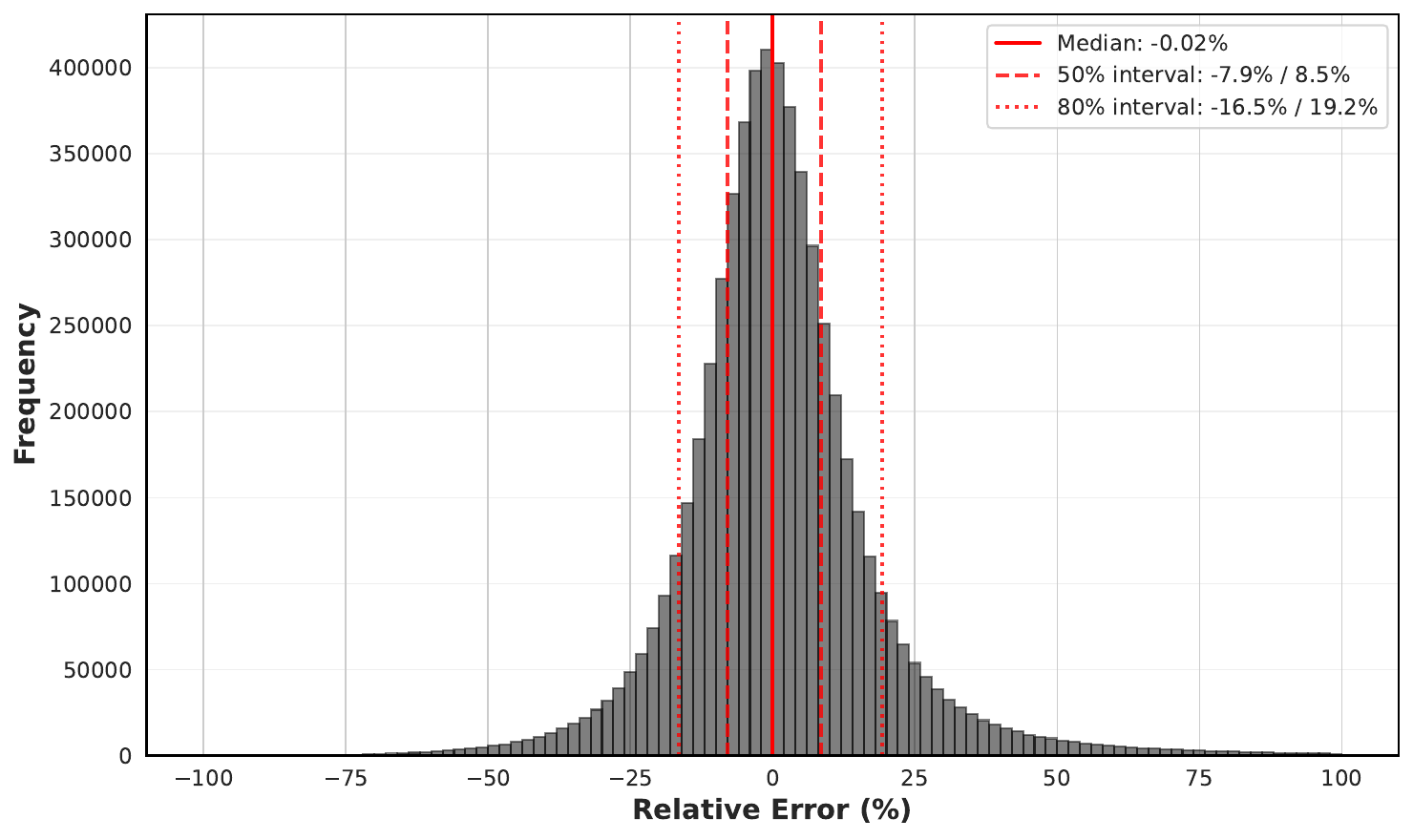}
\caption{{\bf Distribution of relative errors in predictions.} Each bar represents the number of instances for which the relative errors between predicted and actual in percent fall in a specific bin. The median and the intervals containing 50\% and 80\% of the relative errors are shown.}
\label{fig:regression_rel_err}
\end{figure}

Figure~\ref{fig:regression_rel_err} shows the distribution of the relative errors across the whole yield range. It complements the previous graph by showing the actual spread of predictions around the actual values. Around 80\% of the relative errors across all predictions fall between -16.5\% and 19.7\%, demonstrating strong overall performance on the regression task.  

As for the classification task, these two graphs show that the regression model performs well, with notably better results for yields around 100~g to 1 kg TNT.

\subsection{Feature importance and peak ratios selection}

From both machine learning and physics perspectives, it is crucial to identify the most informative features in a dataset, i.e., the peak ratios that the model relies on most for classification or regression tasks. This can not only provide physics interpretability of the model results, but can also help design more efficient inference models relying on reduced set of features.

\subsubsection{Classification}

\begin{figure}[H]
\centering
\includegraphics[width=1.0\linewidth]{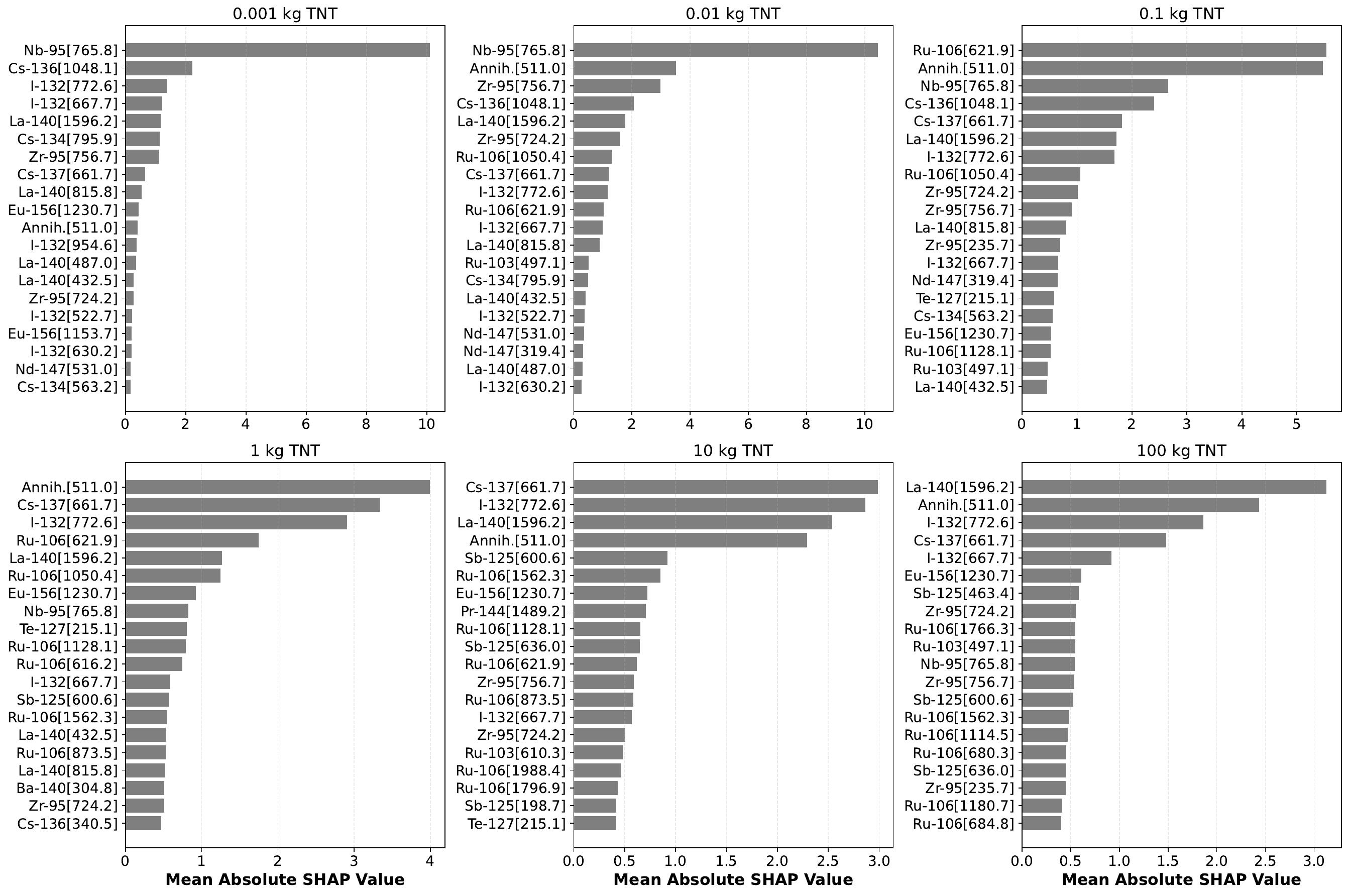}
\caption{{\bf Top 20 peak ratios by SHAP values (classification).}
Each subplot corresponds to a different yield threshold.
A SHAP value of 1 shifts the predicted probability from 50\% to approximately 73\%;
a value of 3 shifts it from 5\% to approximately 95\%.
Features are fission product gamma-ray peak count ratios relative to the 375.1~keV line of plutonium-239.}
\label{fig:top20_SHAP_classification}
\end{figure}

Figure \ref{fig:top20_SHAP_classification} shows the top 20 peak ratios by their mean absolute SHAP value for different yield thresholds. 
Here, the mean absolute SHAP value measures, on average, how strongly a ratio pushes XGBoost's raw output score (which is then converted into a class probability via the sigmoid function), with larger values indicating a stronger influence on the prediction. We can see that a few ratios dominate in importance, especially in the lower-yield range, such as niobium-95 at 765 keV and ruthenium-106 at 621 keV. The 511 keV peak in the gamma spectrum does not originate from gamma emission by a single fission product. It arises from positron-electron annihilation, where the positron was either created by pair production or by a $\beta^+$ decay. This line can be seen as a group effect which holds information on the concentration of fission products in the debris. The importance of these lines lies in their detectability in the considered yield range and their ability to clearly indicate the transition between below- and above-yield thresholds, despite interference from other parameters (time after test, shielding, plutonium mass, and pre-test configuration).

\begin{figure}[H]
\centering
\includegraphics[width=1.0\linewidth]{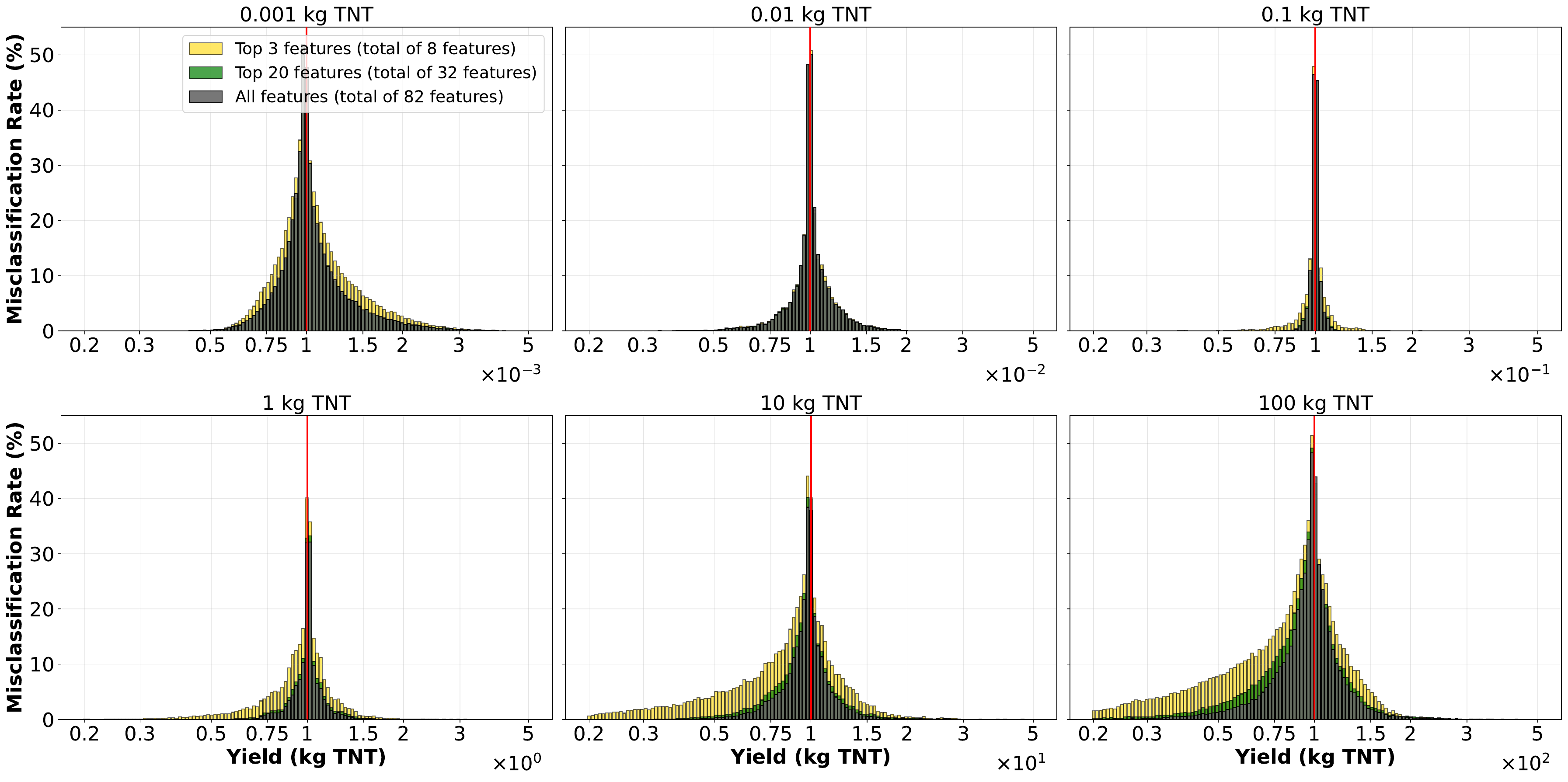}
\caption{{\bf Misclassification rate for various yield thresholds (comparison with reduced models).}
 Each graph shows the distribution of misclassification rates for groups of the same-yield instances around a specific yield threshold. Each same-yield group comprises 13,200 instances. Three sets of data are presented: the original model, the model trained on the top 20 ratios list, and the model trained on the top 3 ratios list.}
\label{fig:misclassification_yield_comparison}
\end{figure}

Given this ranking, it is meaningful to investigate how the model's performance changes as the number of features is reduced while keeping the most important ones. Figure \ref{fig:misclassification_yield_comparison} compares the performance of the original model to two models trained on reduced sets of ratios. One model only uses the ensemble of ratios resulting from merging the top 20 ratios for the six thresholds presented in Figure \ref{fig:top20_SHAP_classification}, while the other model uses a list obtained from merging the top 3 ratios across thresholds (see Appendix \ref{app:S1_Table} for details of these lists). These graphs show that the model's performance on the classification task is virtually the same when limiting the list of ratios to the top 20 list. For the top 3 list, the model becomes less accurate at higher thresholds, but presents similar performance with the original model for lower thresholds (1 g, 10 g, and 100 g), reflecting the predominance of a few ratios in feature importance in this yield range as explained above.

\subsubsection{Regression}

\begin{figure}[H]
\centering
\includegraphics[width=0.5\linewidth]{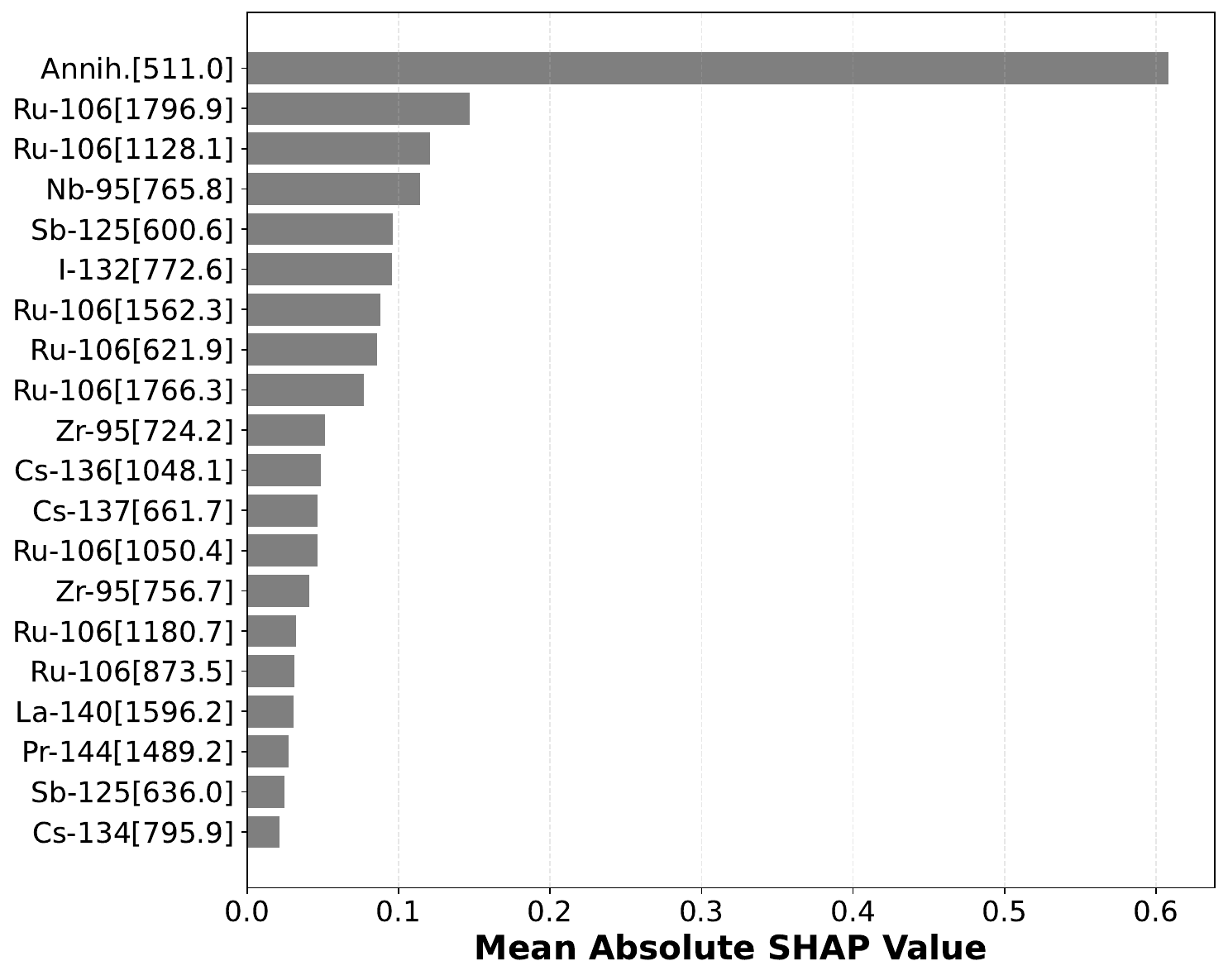}
\caption{{\bf Top 20 peak ratios by SHAP values (regression).} SHAP values are in units of decades of yield: a value of 0.5 indicates that the feature shifts the predicted yield by approximately half an order of magnitude.}
\label{fig:top20_SHAP_classification_regression}
\end{figure}

For the regression task, a single SHAP value ranking has been computed across the whole yield range, as presented in Figure \ref{fig:top20_SHAP_classification_regression}. Here, the mean absolute SHAP value measures, on average, how strongly a feature pushes XGBoost's predicted $\text{log}_{10}$(yield in kg TNT), with larger values indicating a stronger influence on the predicted yield, regardless of direction. The 511.0 keV annihilation line is confirmed to be the predominant feature for the regression task, followed by other lines also observed in the ranking for the classification task.

\begin{figure}[H]
\centering
\includegraphics[width=0.8\linewidth]{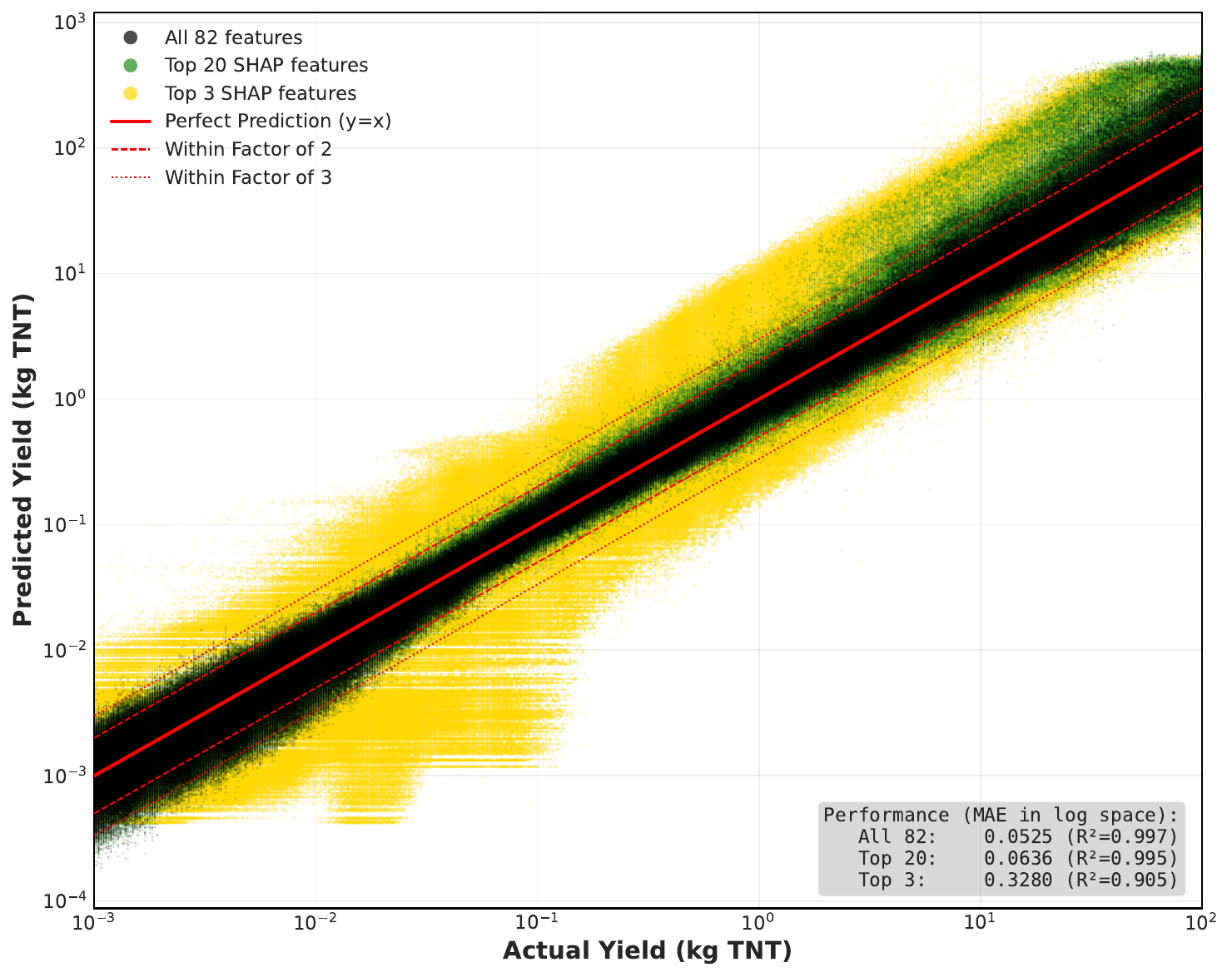}
\caption{{\bf Predicted vs. actual yield values (comparison with reduced models).} Predictions of the model trained with the full set of ratios are plotted as black dots, those of the model trained with the top 20 ratios are in green and those of the model trained with the top 3 ratios are in yellow. The red dashed line represents perfect prediction. The two thinner dashed lines delimit regions where the prediction is within a factor of 2 and 3, respectively, of the actual value. The $R^{2}$ values calculated in $\text{log}_{10}$ space are also shown for the three models.}
\label{fig:regression_predicted_actual_top}
\end{figure}

Figure \ref{fig:regression_predicted_actual_top} shows how the yield predictions change when the model is trained on a reduced set of ratios. This time, the top-3 and top-20 models are simply the models trained with the top 3 and top 20 ratios from Figure \ref{fig:top20_SHAP_classification_regression}, respectively. We see that the top-20 model performs as well as the original model at lower yields but becomes less accurate and precise at higher yields. The top-3 model, which performs poorly overall, shows the opposite trend with better predictions at higher yields and catastrophic predictions at lower yields. These patterns result from the way the top ratios are selected for the regression model: the mean SHAP value of each ratio is the average across the whole yield range. The top-3 list does not include niobium-95, which has the highest feature importance at lower yields.

\subsection{Impact of other parameters on model's performances}

It is important to note that the varying performances of the two models with other parameters are heavily dependent on the processing and selection of features on which the models are trained and tested, and do not necessarily reflect fundamental physical properties or processes unique to these parameters' values (e.g., using a lower mass of plutonium during a test makes the model perform worse not because it is inherently more difficult to predict yields for low plutonium mass values, but because plutonium lines are becoming less detectable and the availability of data for training deteriorates). This aspect can be seen as consequential to the physics-informed approach adopted in this work. The impact of the pre-test configuration of the assembly is not investigated, as only two extreme cases were simulated.

\subsubsection{Classification}

\begin{figure}[H]
\centering
\includegraphics[width=0.8\linewidth]{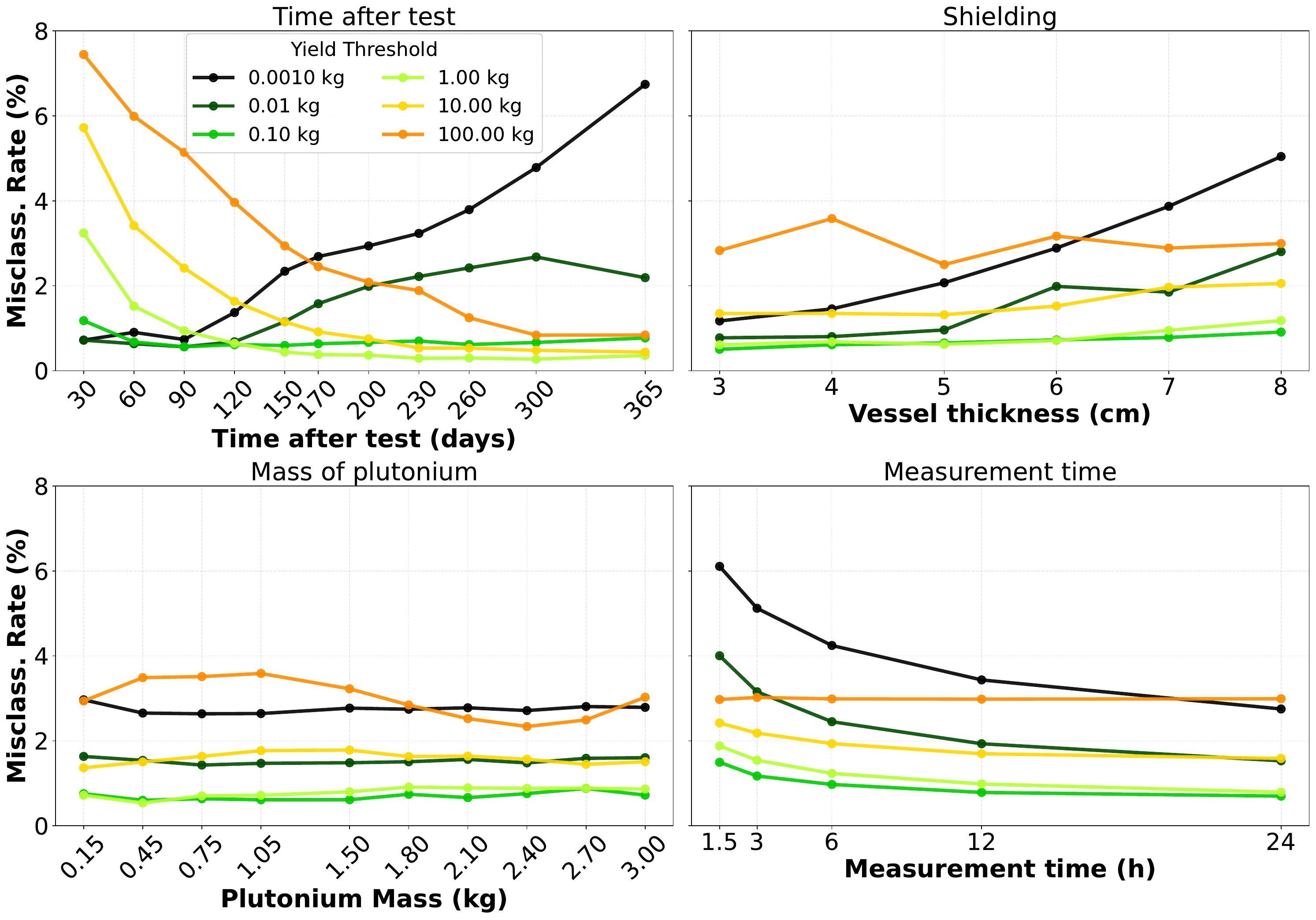}
\caption{{\bf Impacts of parameters on model's performance (classification).} The misclassification rate for instances with the same value of the specified parameter is plotted against these values. One dataset is plotted for each yield threshold.}
\label{fig:class_four}
\end{figure}

Figure \ref{fig:class_four} shows the model's misclassification rates across different parameter values and yield thresholds (one curve per threshold). The upper-left plot, which investigates the impact of the time after test, shows the most complex pattern. At the lowest threshold (0.001 kg TNT), measurements taken one year after the test are more likely to be misclassified than those obtained one month after the test. In contrast, for higher thresholds, measurements acquired a year after the test are less prone to errors. The model is similarly effective across all time intervals for thresholds near 0.1 kg and 0.01 kg TNT, which coincide with the threshold region where the model achieves its best overall performance, as shown in Figure \ref{fig:classification_metrics}. This specific pattern can be explained as follows: tests at lower yields (near 0.001 kg TNT) produce fewer fission products, and their concentration decreases over time due to nuclear decay, leading to detection problems as the time after test increases. For tests at higher yields, plutonium lines are initially buried among fission product lines, but become detectable over time, thereby increasing the quality and quantity of data on which the model is trained.

The upper-right plot shows misclassification rates as a function of shielding effects. Increasing shielding adversely affects the model's performance for lower yields, as it worsens the detectability of already weak fission-product gamma lines. However, this pattern fades away with increasing yield, as the stronger signal overall means that attenuation by the shielding effect (whatever its magnitude) does not impact the detectability of lines (fission products or plutonium).

The lower-left plot presents the misclassification rates against the mass of plutonium. While the model seems unaffected by the mass for most yields, it presents non-monotonic patterns for the two higher thresholds. This behavior arises because the peak ratios implicitly encode plutonium mass through the denominator, causing the classifier to partially learn mass-specific decision branches rather than purely yield-discriminative rules. This aspect is particularly pronounced for high yields because these are instances where spectral data filtering based on plutonium-239 detectability affects the training the most. More analysis on these results can be found in Appendix \ref{app:S2_Text}.

Finally, the lower-right plot shows the impact of measurement time of the gamma spectrum. As expected, the model's performance improves with longer measurement times, as more fission product lines become detectable and the signal-to-noise ratio improves. This improvement diminishes continuously as the yield threshold increases, becoming negligible at 100~kg TNT. In the vicinity of this threshold, all selected yields produce fission-product signals well above the detection limit even at 1.5~hours, so the feature vectors are effectively saturated and longer counting provides no additional discriminative information.

\subsubsection{Regression}

\begin{figure}[H]
\centering
\includegraphics[width=0.8\linewidth]{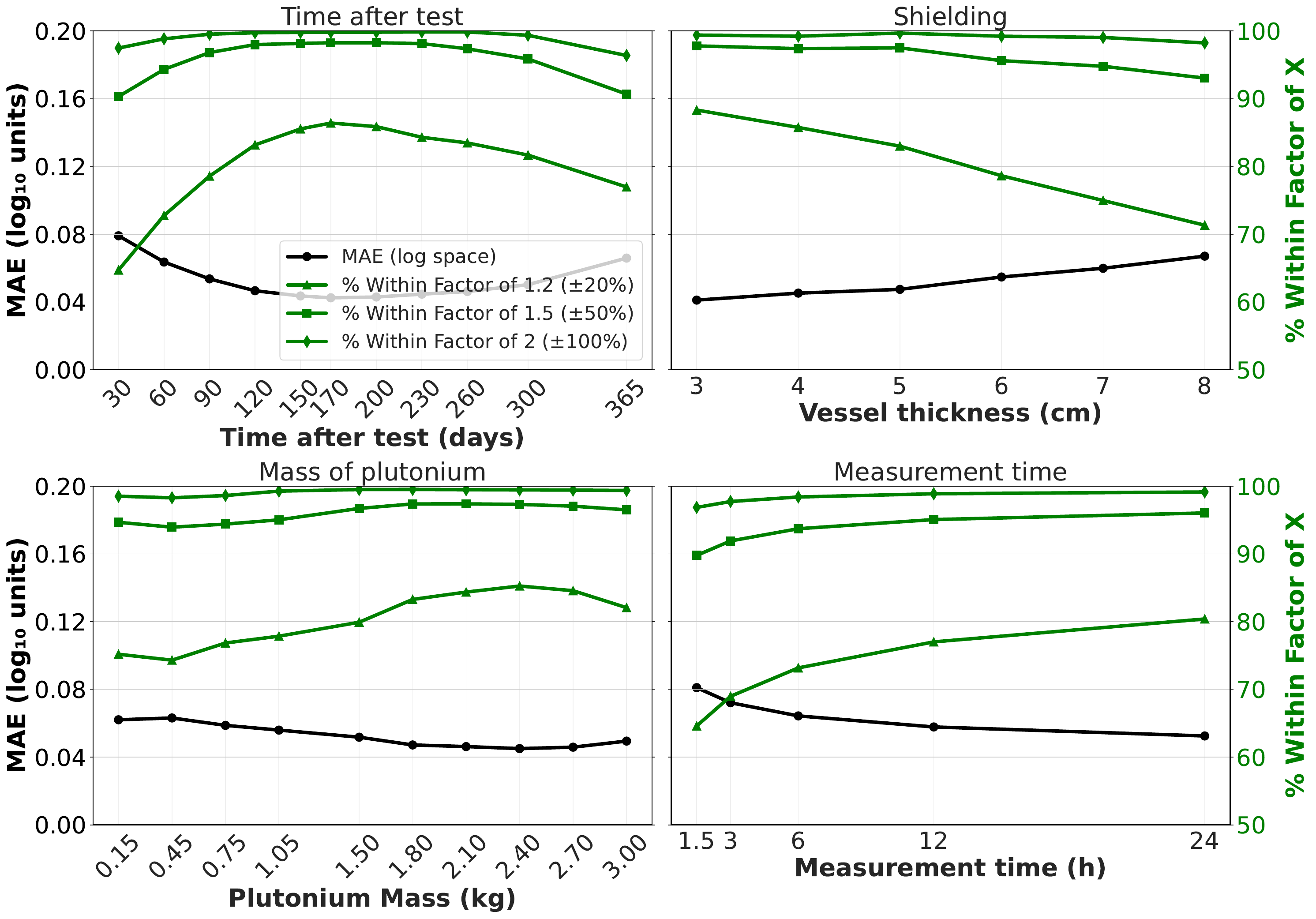}
\caption{{\bf Impacts of parameters on model's performance (regression).} The log10 MAE and the \% within 1.2, 1.5, and 2 factors for instances with the same value of the specified parameter are plotted against these values.}
\label{fig:regress_four}
\end{figure}

Similar results are obtained for the regression model, as can be seen in Figure \ref{fig:regress_four}. Here, metrics such as the \(\log_{10}\text{MAE}\) and the percentage within a factor of x represent yield-average metrics, contrasting with the per-yield-threshold metrics presented for the classification model. The top-left plot shows the impact of the time after test and indicates that the model performs better in the middle time range, whereas performance worsens when the spectrum is measured earlier or later. For the impact of shielding effects, the regression model's performance decreases with increasing shielding, as shown in the top-right plot. The bottom-left plot shows that the mass of plutonium carries similar effects on the regression as for the classification model. Finally, as expected, longer measurement times improve the regression model, as shown in the bottom-right plot.

\subsection{Training sensitivity and strategies}

Beyond demonstrating the performance of a given model for the classification and regression tasks, it is also important to investigate how the training impacts the model. Since a deployed model for very low-yield test verification would have to rely on sets of simulated data for training, the developers have a certain control over these datasets. Key considerations, such as the optimal ranges for test and measurement parameters or the value of incorporating supplementary features beyond gamma peak ratios, must be addressed to maximize the model's efficiency and efficacy.

\subsubsection{Parameters' training ranges}

\begin{figure}[H]
\centering
\includegraphics[width=0.9\linewidth]{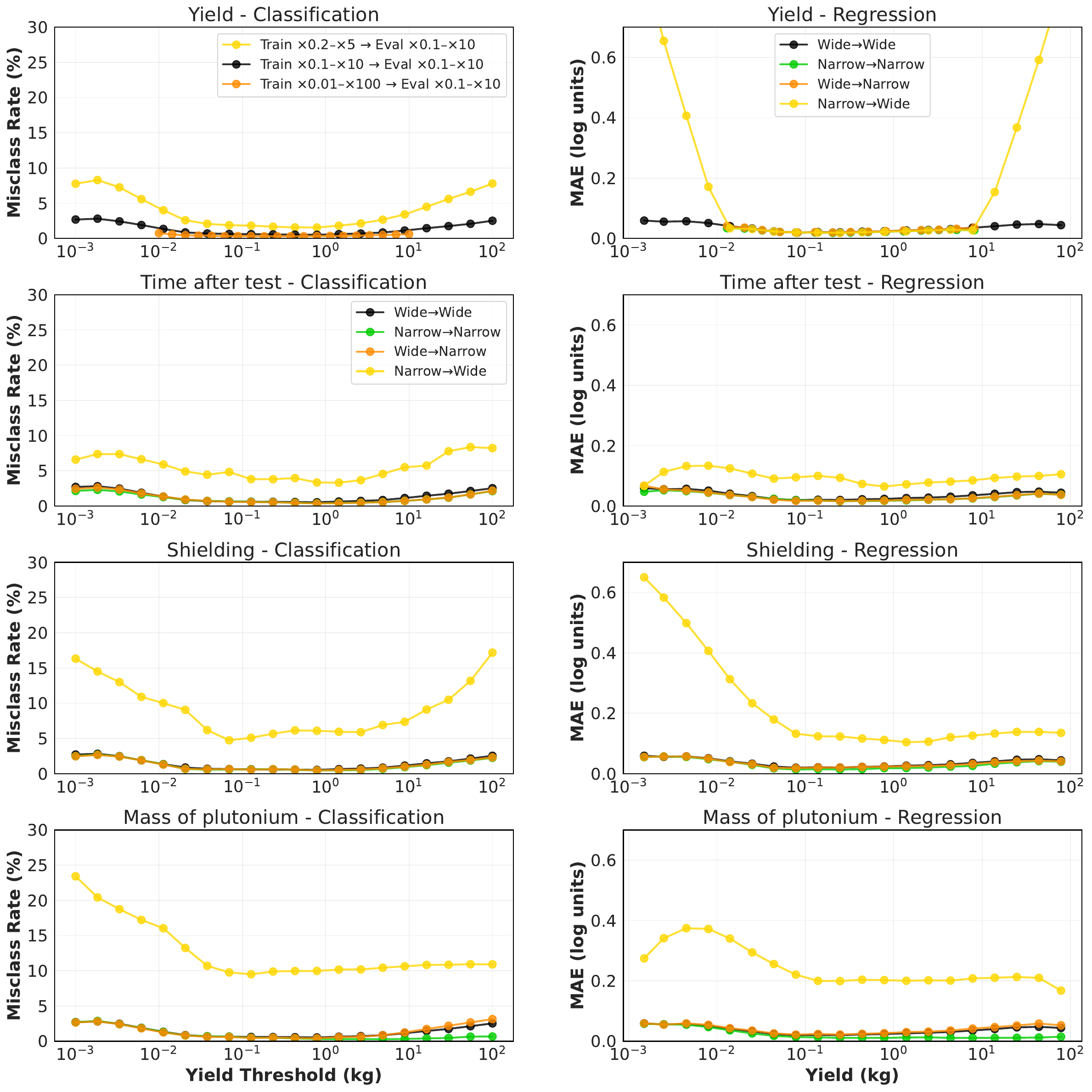}
\caption{{\bf Performance of classification and regression models under different training range strategies.} The classification performance is displayed with misclassification rate (\%) and the regression performance with the log10 of the MAE. For both tasks, four different training-testing range combinations are investigated.}
\label{fig:training_range}
\end{figure}

Figure \ref{fig:training_range} shows how the training ranges used for test and measurement parameters impact the performance of the model, both for classification and regression tasks, using respective metrics. To investigate the impact of the yield training range on the classification task, we used three different threshold-specific windows for the training, while the evaluation window remained the same (top-left subplot). For all other subplots, two ranges are used for training and testing: a narrow and a wide range (Appendix \ref{app:S2_Table} provides details on these ranges). Figure \ref{fig:training_range} presents how different combinations of these ranges in the training and the testing of the models impact their performance. For instance, curves labeled ``wide-narrow'' mean that the model was trained on the wide range and tested on the narrow range for the corresponding parameter. When one parameter is tested, all other parameters are set to their wide range for the training and testing (and to $\times 0.1 - \times 10$ for the yield in classification tasks). 

Focusing on the top-left plot, we see that the model trained on the narrower window presents the poorest performance. This indicates that the model struggles to classify instances that lie beyond its training range. On the other hand, we see that the model trained on a wider window does not diverge too much from the default model. This means that training on a wider window neither penalizes nor significantly improves the classifier.

For the remaining subplots, across all parameters and for both tasks, the narrow-wide combination shows the poorest results. Similarly to the top-left subplot, this indicates that the model struggles to extrapolate patterns across regimes beyond its training range. On the "Yield - Regression" subplot, we can clearly see that the model makes accurate predictions on its training range but starts to severely diverge outside of it. While the drop in performance when training on a narrower time range seems to be uniform across yields, we observe yield-dependent patterns when the model is trained on narrower shielding and plutonium mass ranges.

We also observe that the curves associated with wide-narrow and narrow-narrow combinations overlap for the yield (regression), the time, and the shielding parameters. This implies that training on wider ranges for these parameters does not improve nor hurt the performance of the model. However, we see that the narrow-narrow model outperforms the wide-narrow model for the plutonium mass in the high-yield regions. This means that using a wider-than-necessary plutonium mass range for training hurts the model for classifying or predicting in the high-yield range. 


Overall, these results confirm the confounding nature of the three parameters (time after test, shielding, and plutonium mass) while guiding optimal strategies for training machine learning models. Specifically, they indicate that although narrowing the training range for certain parameters based on evidence (e.g., plutonium mass) can offer benefits, the drawbacks of overly restrictive ranges outweigh these advantages. Consequently, this work advocates a conservative approach favoring broader training ranges.

\subsubsection{Adding parameters as training features}

\begin{figure}[H]
\centering
\includegraphics[width=0.9\linewidth]{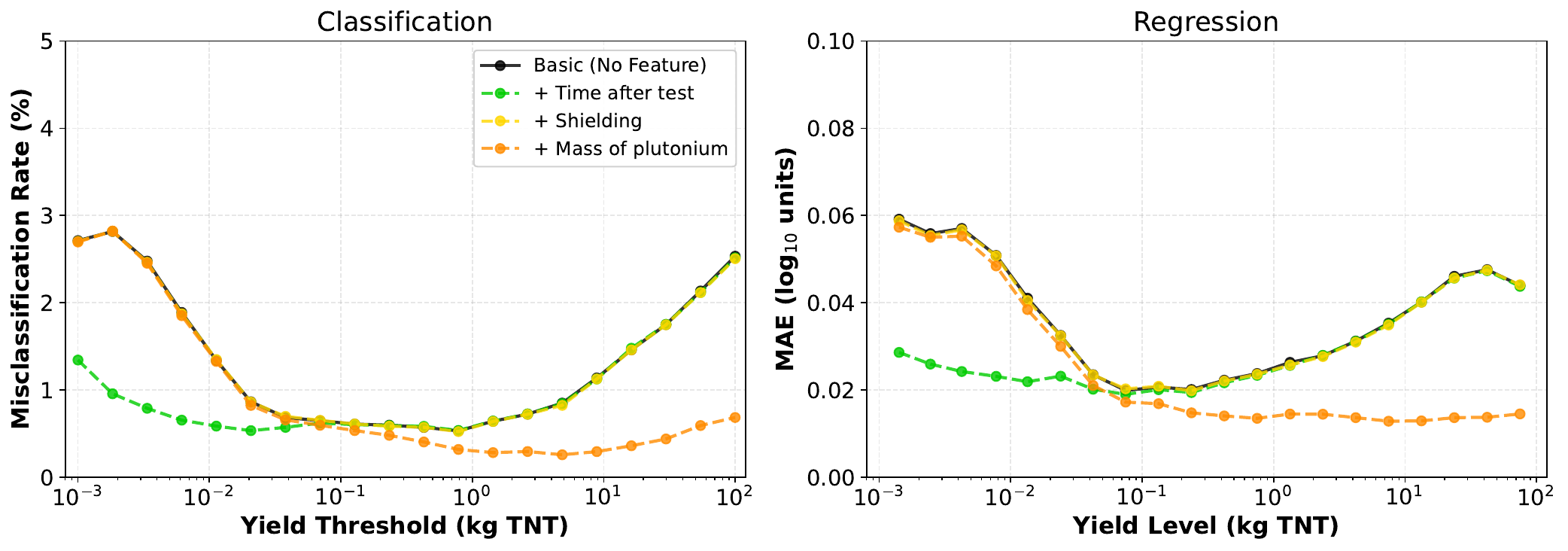}
\caption{{\bf Performance of classification and regression models under different input feature strategies.} The classification performance is displayed with misclassification rate (\%) and the regression performance with the log10 of the MAE. For both tasks, the performance of the basic model is compared with the augmented model where the indicated parameter is added as an input feature.}
\label{fig:feature_impact}
\end{figure}

In previous approaches, the peak ratios from the spectra were the only features used to train and test the model. Here, we investigate the model's performance when parameters such as the time after test, the shielding effect, and the mass of plutonium are added as input features. Figure \ref{fig:feature_impact} plots the respective metrics for the classification and regression tasks, comparing the basic model and the model augmented by adding the indicated parameter as an input feature. We observe no clear improvement when adding the shielding effect (via the variable vessel thickness) as an input feature. However, the model performs better in the lower yield ranges when the time after test is used as an input feature. Similarly, the model performs better in the higher-yield region when plutonium mass is added as an input feature.

These results show that adding specific parameters as input features can improve model performance, especially in specific yield regions. However, this strategy is double-edged, as it requires inspectors to have high confidence in the parameter's real-world value for the very low-yield test being verified. If the inspectors' estimate for this parameter is inaccurate, the augmented model will perform worse than the basic model.

\section{Discussion}

Overall, the results of this work demonstrate that ML methods have significant potential for inferring the fission yield level of very low-yield tests using gamma spectroscopy of debris remaining after the test. Specifically, methods based on decision trees (XGBoost, Random Forest) achieve high performance for classifying and predicting yield values. These results can lay the foundation of further research and efforts to develop, test, and deploy effective verification technology for very low-yield tests. However, these efforts will have to be embedded into and inform the policy and geopolitical environment surrounding the verification of very low-yield tests.

The current accepted standard for assessing the compliance of very low-yield tests with the principle of "zero yield" is the zero-yield standard, which imposes a limit on the criticality level (no supercritical fission chain reaction) rather than on the production of a fission yield. Based on our research, we understand that gamma spectroscopy provides only limited insight into the criticality level of a past test. However, more insights can be drawn if the verifying party obtains information on certain other factors. Obtaining independent information on the yield level, for instance, can help narrow the estimate of the criticality level, as the two terms are related, as shown in equation \ref{eq:yield}. A regression model such as the one presented in this work could therefore be used as part of verification processes to infer the criticality level with post-test gamma spectroscopy. Yet, such an approach will only offer limited estimation on the range of the criticality level, which might not, ultimately, allow for making final conclusions on the compliance of a very low-yield test with the zero-yield standard.

A yield threshold-based standard might be much easier to verify and more effective at diffusing tensions and preserving a nuclear test ban. An arms control standard designed for compliance verification should address two (sometimes conflicting) objectives: achieving the ultimate goals of the effort or treaty in which it is used, and enabling practical, effective verification that permits drawing a conclusion on compliance with a satisfying level of confidence for all parties. It can be argued that the current zero-yield standard provides neither. On the other hand, this work shows that ML methods can effectively classify yield levels compared to a given threshold in relevant yield ranges. A potential yield threshold-based standard could therefore be more effectively verified than the zero-yield standard. Whether such a yield threshold-based standard would achieve the ultimate objectives of a nuclear test ban effort or treaty depends on these objectives and on the chosen threshold level. We are fairly confident that a threshold set at around 1 g TNT would render a very low-yield test unusable for developing new warhead designs for actors with limited knowledge of nuclear weapons science. However, it is unclear to what extent the same threshold could limit modernization programs and arms races between nuclear-weapon states. A lower yield threshold (in the mg range) might be more desirable but could be more challenging to verify.

History shows that such verification gaps can be progressively closed. The 1974 Threshold Test Ban Treaty capped underground weapon tests at 150 kt but left peaceful nuclear explosions (PNEs) as a loophole; the Peaceful Nuclear Explosions Treaty, signed two years later, extended the same limit to PNEs and introduced on-site-inspection provisions. Similarly, post-1991 revelations about Iraq’s clandestine nuclear program exposed the limits of traditional IAEA safeguards (which covered only declared facilities), prompting the 1997 Additional Protocol that expanded inspector access to undeclared sites and activities. These precedents suggest that identifying a verification gap is often the first step toward closing it, and that the current limitations for very low-yield tests are amenable to technical and procedural solutions rather than an immutable feature of the regime.

It is important to note that a spectroscopy-based verification technique like the one discussed here would have to be embedded in a broader verification procedure which would include collections of information that can improve and enrich the spectroscopy-based verification. For instance, satellite imagery analysis can help narrow down possible time intervals between test and measurement by identifying preparation activities at the testing site. As shown in this work, such information can guide ML model training to improve yield inference performance. Transparency measures introduced in an agreed-upon treaty framework could allow inspectors to collect and learn various information that will also improve the spectroscopy-based verification technique (i.e., mass of plutonium used, containment vessel material, and thickness).

The training strategies of the ML models are ultimately the sole aspect of this verification method over which the verifying party has full control. We do not expect experimental data from very low-yield tests to be readily available for training ML models. Physics model simulations of very low-yield tests are computationally intensive but can be run iteratively and accumulated over time to produce large datasets of simulated spectra that will train ML models. When verification must be conducted, ML models can be trained on different slices and subspaces of these datasets, tailored to the specifics of that verification event. The measured data could be registered and archived, allowing further evaluation, and could also serve as a reference standard for real-world data that could enrich the dataset. These models and datasets could be made open source to allow different communities to train new ML models and test alternative techniques for inferring fission yields.

\section{Conclusion}

This work presents, to our knowledge, the first efforts to apply machine learning methods to infer the fission yield of very low-yield tests (here defined as any nuclear explosive test producing less than 1 ton TNT together with subcritical tests) from spectroscopy measurements of radioactive debris. Our approach relies on training machine learning models on computed gamma peak ratios extracted from 66 million gamma spectra simulated using high-fidelity physics models of very low-yield tests across a wide range of test and measurement parameters. 

Overall, our results show that decision tree-based methods perform best both for classifying whether the yield is below or above a set threshold and for estimating the actual yield value. Through feature importance analysis, we identify the most important peak ratios that help ML models make decisions. We also investigate the impact of other confounding parameters, such as the time after test, shielding effects, and plutonium mass, on the ML models' performance, providing useful information to optimize their training.

Future work should explore Bayesian models to further improve ML performance when information about yield or confounding parameters becomes available and explore technics such as conformal prediction for uncertainty quantification. If experimental data becomes available (even in small quantity) transfer learning from simulations to real data and domain adaptation can be used to improve the reliability and generalization of the method. In addition, while this work adopted a ``physics-informed'' approach for the training of the ML models (by solely using gamma peak ratios and filtering those that were undetectable), future work can explore a more unconstrained training approach, for instance, by including all the raw spectra counts in the training or by specifically focusing on deep learning models.

\section*{Acknowledgments}
The authors would like to thank the Academy of Interdisciplinary Studies at HKUST for providing funding for this research.

\section*{Data availability statement}

The peak-ratio dataset corresponding to spectra measured for 24 hours (13.2 million spectra) is deposited on figshare (doi:10.6084/m9.figshare.32118838).
The rest of the dataset (spectra measured for 1.5, 3, 6, and 12 hours) was not posted due to data size limitations on figshare, but can be provided upon special request.
All code for gamma-spectrum simulations and machine learning is open source and can be found on GitHub: \\
\noindent OpenMC: \url{https://github.com/openmc-dev/openmc} \\
\noindent ONIX: \url{https://github.com/jlanversin/ONIX} \\
\noindent decaypy: \url{https://github.com/cfichtlscherer/decaypy} \\
\noindent XGBoost: \url{https://github.com/dmlc/xgboost}

\section*{Author contributions statement}

J.d.T.d.L: Conceptualization; Formal Analysis; Funding Acquisition; Investigation; Methodology; Resources; Software; Supervision; Validation; Visualization; Writing – Original Draft, Review, and Editing. J.L: Data Curation; Formal Analysis; Investigation; Software; Validation. C.F.: Data Curation; Methodology; Writing - Review, and Editing. D.S: Methodology, Writing - Review. M.K.: Conceptualization; Writing - Review.
All authors reviewed the manuscript.

\section*{Competing interest statement}

The authors declare no competing interests.

\appendix

\section{Models architectures and settings}
\label{app:S1_Text}
The multilayer perceptron (MLP) consisted of two hidden layers with 100 neurons each, tanh activation functions, a stochastic gradient descent (SGD) optimizer with an adaptive learning-rate schedule (initial value 0.01), and L2 regularization with penalty 0.0001. For the support vector classifier (SVC) and k-nearest neighbors (KNN) models, input features were standardized using \texttt{StandardScaler} from scikit-learn. No feature scaling was applied to the tree-based models (XGBoost, Random Forest, and Decision Tree).

\section{Variation of data availability with yield}
\label{app:S1_Figure}

\begin{figure}[H]
\centering
\includegraphics[width=0.7\linewidth]{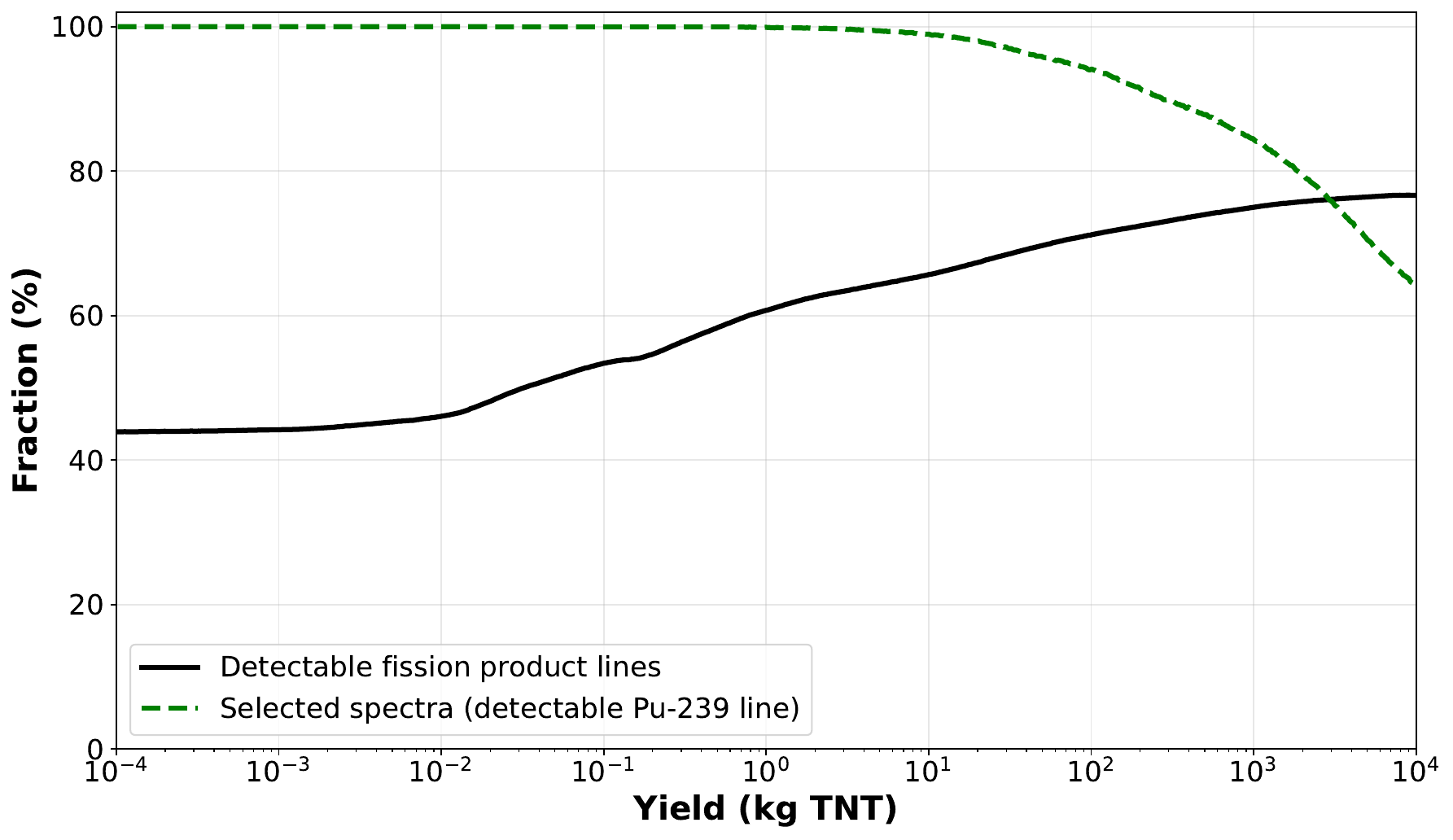}
\caption{{\bf Variation of data availability with yield value.} The dashed green curve shows the fraction of selected spectra over the total generated spectra for a given yield value. The plain black curve shows the fraction of non-zero peak ratios (detectable fission product line) averaged over all selected spectra of a same yield level. At low yields, fission products concentration are low so fission products line are less intense. At high yields, fission product lines become more intense and the plutonium line becomes buried by them.}
\label{fig:S1_figure}
\end{figure}

\section{Top SHAP features selected for binary classification models}
\label{app:S1_Table}

\begin{table}[H]
\caption{Top SHAP features selected for binary classification models. 
The numbers in parentheses indicate the total gamma lines and unique isotopes after merging top features across all six yield thresholds (0.001, 0.01, 0.1, 1, 10, 100 kg TNT). Energies are in keV.}
\begin{tabular}{l| p{9.5cm}}
\toprule
\textbf{Feature Set} & \textbf{Gamma Lines (grouped by isotope)} \\
\thickhline
Top 3 SHAP (8 lines, 8 isotopes) & 
Annih.[511.0]; Cs-136[1048.1]; Cs-137[661.7]; I-132[772.6]; La-140[1596.2]; Nb-95[765.8]; Ru-106[621.9]; Zr-95[756.7] \\
\midrule
Top 20 SHAP (45 lines, 16 isotopes) & 
Annih.[511.0]; Ba-140[304.8]; Cs-134[563.2, 795.9]; Cs-136[340.5, 1048.1]; Cs-137[661.7]; Eu-156[1153.7, 1230.7]; 
I-132[522.7, 630.2, 667.7, 772.6, 954.6]; La-140[432.5, 487.0, 815.8, 1596.2]; Nb-95[765.8]; Nd-147[319.4, 531.0]; 
Pr-144[1489.2]; Ru-103[497.1, 610.3]; Ru-106[616.2, 621.9, 680.3, 684.8, 873.5, 1050.4, 1114.5, 1128.1, 1180.7, 1562.3, 1766.3, 1796.9, 1988.4]; 
Sb-125[198.7, 463.4, 600.6, 636.0]; Te-127[215.1]; Zr-95[235.7, 724.2, 756.7] \\
\bottomrule
\end{tabular}
\label{tab:top_features}
\end{table}

\section{Detailed analysis of misclassification rate variation with plutonium mass}
\label{app:S2_Text}
For high yield thresholds, the classifier exhibits a non-monotonic dependence of misclassification rate on plutonium mass. This behavior originates from the feature representation: because the 82 classification features are ratios between fission product lines and a selected plutonium-239 line, the magnitude of every feature implicitly encodes mass. When training on all masses simultaneously, the tree-based classifier exploits this mass-dependent scaling to construct mass-specific decision branches rather than learning purely yield-discriminative rules. The decrease in misclassification rate toward the lower mass boundary (0.15~kg) occurs because feature magnitudes at this mass are approximately 20 times larger than at higher masses, making these samples so distinct in feature space that the classifier can route them into dedicated decision branches with minimal interference from other mass groups. The decrease from the peak at 1.05~kg to the minimum at 2.40~kg is explained by two reinforcing effects: first, improved data availability at higher masses (detectability censoring removes 25\% of samples at 0.15~kg but only 2.6\% at 3.0~kg), and second, reduced inter-mass interference, as adjacent high-mass feature distributions become increasingly similar, allowing the globally-learned decision boundary to serve multiple masses without conflicting rules. At 3.00~kg, this advantage is lost because the mass lies at the edge of the training range with only one close neighbor, and the decision boundary is pulled inward by the influence of lower-mass training data, degrading accuracy. Two independent tests confirm this interpretation. First, per-mass locally-trained models (with identical architecture and cross-validation scheme) yield a monotonically decreasing error curve, consistent with the expected effects of reduced data availability at lower masses. Second, applying per-mass $z$-score normalization to the features before global training removes the mass-encoded magnitude information and recovers a monotonically decreasing curve closely matching the local models. Because plutonium mass is not known \textit{a priori} in a realistic deployment scenario, the results presented in the main text use the unnormalized global model, and the non-monotonic pattern should be understood as a consequence of the feature representation rather than an intrinsic property of the classification task.

\section{Ranges used for training sensitivity analysis}
\label{app:S2_Table}

\begin{table}[H]
\caption{Narrow and wide parameter ranges used in the training strategy sensitivity analysis.}
\begin{tabular}{l|c|c}

\textbf{Parameter} & \textbf{Narrow range} & \textbf{Wide range} \\
\thickhline
Yield (kg TNT) & 0.01--10 & 0.001--100 \\
Time btw test and measurement (days) & 90--260 & 30--365 \\
Vessel thickness (cm) & 5--6 & 3--8 \\
Plutonium mass (kg) & 1.05--2.10 & 0.15--3.00 \\
\end{tabular}
\label{tab:param_ranges}
\end{table}


\end{document}